\pdfoutput=1
\documentclass[11pt,a4paper]{article}
\usepackage{jheppub}
\usepackage[T1]{fontenc}
\usepackage{lmodern,amsthm,mathtools,bm,mathrsfs,booktabs,tabularx,array,microtype}
\usepackage{enumitem,needspace,xcolor,placeins}

\setlist{nosep,leftmargin=1.8em}
\renewcommand{\Re}{\operatorname{Re}}
\renewcommand{\Im}{\operatorname{Im}}
\newcolumntype{Y}{>{\raggedright\arraybackslash}X}
\numberwithin{equation}{section}

\newcommand{\dd}{\mathrm d}
\newcommand{\GN}{G_N}
\newcommand{\OO}{\mathcal O}
\newcommand{\PP}{\mathcal P}
\newcommand{\sech}{\operatorname{sech}}
\newcommand{\arctanh}{\operatorname{arctanh}}
\newcommand{\Var}{\operatorname{Var}}
\newcommand{\norm}[1]{\left\lVert#1\right\rVert}

\title{Modular tomography of perturbative quantum causal shadows II: Local positivity}

\author[a]{Liangyu Chen} 
\affiliation[a]{Yau Mathematical Sciences Center, Tsinghua University, Beijing 100084, China} 

\emailAdd{liangyu-chen@mail.tsinghua.edu.cn}

\hypersetup{pdftitle={Modular tomography of perturbative quantum causal shadows II: Local positivity},pdfauthor={Liangyu Chen}}
\abstract{In vacuum AdS$_3$/CFT$_2$, Paper I related a positive boundary modular response to a radial average of the displacement of a quantum extremal surface (QES) probed by a single Euclidean scalar insertion. Here we upgrade that averaged inequality to exact local positivity using only the same scalar species, and apply the resulting tomography to the scalar QES. Scanning the scale of one boundary insertion determines the proper displacement profile $H$ from ideal data and directly gives the leading spatial area of the causal shadow, yet an explicit smooth counterexample shows that strictly positive averages at every scale can conceal a localized negative displacement. In the same tree-level Einstein regime, we extend Paper I's complete-graph $AA$ and cross-channel remainder bounds to finite coherent superpositions of Euclidean preparations, including horizon-flux interference. Finite preparation derivatives act as commuting polynomial multipliers along the compactified horizon, generating normalized states that sample $H$ with weights proportional to $|p|^2$ for arbitrary polynomials $p$. For any bounded profile, positivity of all finite moment matrices at a single scan center is equivalent to $H\ge 0$ almost everywhere. In the ordered limit, this upgrades boundary relative-entropy monotonicity to local nonnegativity and enforces a rigidity saturation criterion, while a shifted finite boost window suppresses both tree-level remainders by powers of $G_N$. For the scalar QES, the entropy source, gravitational propagation, and probe averaging factorize, yielding exact area and central-thickness benchmarks and sharp deconvolution-free pointwise bounds on $H$.}

\begin{document}
\makeatletter
\let\MT@oldtoc\tableofcontents
\renewcommand\tableofcontents{{\linespread{1.0}\selectfont\MT@oldtoc}}
\makeatother
\maketitle
\flushbottom

\section{Introduction}
\label{sec:intro}

This is the second of two papers on the boundary characterization of perturbative quantum causal shadows. The geometric setting follows from the Ryu--Takayanagi prescription and its covariant extension~\cite{RT,HRT}, whose bulk entanglement corrections lead to the quantum extremal surface (QES) description~\cite{FLM,EW}. When the classical extremal and causal surfaces coincide, correlations between disconnected bulk wedges can generate a quantum separation between them. This regime was used by Levine, Shahbazi-Moghaddam and Soni to study how cocycle flow can bring a causally inaccessible part of the entanglement wedge into causal contact with the boundary~\cite{SeeingEW}. Chen and Wang then computed the correlation-induced QES displacement from bulk mutual information and asked how its positive sign is encoded in the boundary conformal field theory (CFT)~\cite{Shape,CW}. Motivated by this question, boundary modular-flow entropy positivity for two disjoint intervals was recently analyzed in ref.~\cite{MES}; our series instead connects boundary relative-entropy monotonicity directly to the bulk QES displacement via high-boost modular scattering. Here our aim is to upgrade that boundary characterization to exact local positivity and apply it to the scalar QES geometry.

The reconstruction of local bulk operators from boundary data~\cite{HKLL} provides the starting point for probing this geometry. Faulkner and Lewkowycz related reconstruction in the entanglement wedge, including operators on the extremal surface, to boundary modular flow~\cite{FL}; intersecting modular Hamiltonians and explicit modular reconstruction kernels developed this approach further~\cite{KLIntersect,FLWToolkit}. Boundary information metrics provide another means of extracting entanglement-wedge geometry~\cite{InfoMetricEW}. For the displacement considered here, the direct dynamical input is the work of Chandrasekaran, Faulkner and Levine, who used high-boost modular correlators to probe QES locations through gravitational scattering~\cite{CFL}, building on the AdS eikonal description of boundary four-point functions~\cite{EikonalAdS}. These constructions motivate a sharp question: how can normalizable probes of a single boundary scalar species constrain the local sign and shape of the displaced surface?

Paper I~\cite{CompanionPositivity} supplies the boundary observable and single-insertion dynamical dictionary whose resolving power we study here. Its boundary sign comes from relative-entropy monotonicity and the associated full-modular inequalities~\cite{CasiniRE,BCFull,ModEnergy,FLPW}. For any fixed single-insertion Euclidean probe in the selected Einstein regime, Paper I showed that the growing cross-channel response measures a horizon-flux radial average of the QES displacement, while both the complete same-side $AA$ contribution (with both light insertions in the same boundary component) and the boost-rescaled integrated cross-channel Regge remainder decay as $O(e^{-\beta})$ and have zero Laplace residue. The positive Laplace projection therefore yields an all-scale averaged positivity constraint on the QES displacement.

A normalizable Euclidean boundary preparation does not sample a single bulk radial point. It prepares a wavepacket of finite transverse width, and high-boost gravitational scattering pairs the QES displacement with the packet's horizon null-energy flux. We work in vacuum $\text{AdS}_3/\text{CFT}_2$, with a half-line and a distant small interval in the disconnected entanglement-wedge phase. If $z$ is the Poincar\'e radius and $\lambda$ the interval size in units of the separation, the selected QES branch is displaced by $\eta(z)=G_N\lambda^{2\Delta}zH(\log z)+\cdots$. Here $G_N$ is Newton's constant, $\Delta$ is the dimension of the light scalar responsible for the correlation, and $H$ is the leading proper-displacement coefficient along the reference curve. A single-insertion probe primary of fixed dimension $d>2$, prepared at scale $s=e^y$, measures the normalized convolution
\begin{equation}
 G_d(y)=\int_{\mathbb R}P_d(y-x)H(x)\dd x,
 \qquad x=\log z,\quad y=\log s,
 \label{eq:introtransform}
\end{equation}
against a strictly positive, unit-normalized kernel $P_d(r)\propto(2\cosh r)^{-2d-1}$. Changing $y$ translates the sampling window; changing the preparation angle rescales the energy prefactor without narrowing the window.

The distinction between an average and a local value is consequential. Exact scale scans determine $H$ uniquely, but their Fourier inversion exponentially amplifies short-distance errors. More importantly for boundary inequalities, positive single-insertion averages at every scale can still conceal a localized negative neighborhood of $H$; we exhibit an explicit smooth profile with a negative central notch whose single-insertion readout is strictly positive at all $y\in\mathbb R$. Recovering a function from exact averages and determining its pointwise sign are therefore distinct problems. By contrast, the integral of the scan requires no deconvolution: it directly measures the leading spatial area of the selected causal shadow.

To gain control over the sampling profile without sacrificing boundary positivity---differentiating the probe state inside the boundary Hilbert space rather than post-processing measured scan data---we upgrade Paper I's single-insertion observable to finite coherent superpositions of distinct Euclidean preparations of the same scalar species, retaining both their off-diagonal horizon interference fluxes and their physical boundary Gram overlaps. Extending Paper I's complete-graph $AA$ and Mellin cross-channel remainder estimates to off-diagonal preparation pairs (Section~\ref{sec:AAinput} and Appendix~\ref{app:offdiag}) transfers exact boundary positivity to the resulting geometric quadratic form on every fixed finite probe subspace. Marolf, Parrikar, Rabideau, Izadi Rad and Van Raamsdonk constructed Lorentzian bulk initial data from Euclidean boundary sources and examined the source cost of localization~\cite{EuclideanSources}. Belin, Lewkowycz and S\'arosi related the geometry of these source-prepared state overlaps to the bulk symplectic form~\cite{BoundarySymplectic}. Our construction uses finite derivatives of a single scalar insertion within this Euclidean preparation framework. They act as commuting polynomial multipliers in a double-Laplace representation of the horizon wavefunction. After the null momentum is integrated out, the response is proportional to
\begin{equation}
 \int_{-1}^1H(y+\arctanh t)|p(t)|^2\dd\mu_d(t),
 \qquad \dd\mu_d(t)\propto(1-t^2)^{d-1/2}\dd t,
 \label{eq:introtest}
\end{equation}
where $p$ is any polynomial of finite degree and $t=\tanh(x-y)$ compactifies the proper offset. The physical boundary norm has a different weight, proportional to $(1-t^2)^{d-1}$. Keeping both measures ensures that every sampling profile corresponds to a normalized state of the original scalar species.

These preparations furnish a complete hierarchy of local positivity criteria. The use of polynomial probability weights and moment matrices is related to measure-based polynomial optimization~\cite{Lasserre2011,PolyOpt}; here the weights are realized by physical boundary states of the original scalar species, with their Hilbert-space norms retained. In an orthogonal-polynomial basis, positivity of every finite moment matrix at a single scan center is equivalent to $H\ge0$ almost everywhere (a.e.) for an arbitrary bounded profile $H\in L^\infty(\mathbb R)$, and pointwise for continuous profiles. Any negative set of positive measure is therefore witnessed at a finite derivative order, allowing the ordered perturbative and boost limits to be taken on a fixed finite superposition subspace before removing the finite-difference step.

It is helpful to distinguish two layers in our results. The scan analysis and moment-matrix completeness relation of Sections~\ref{sec:inverse}--\ref{sec:superposition} are structural statements valid for an arbitrary bounded profile $H\in L^\infty(\mathbb R)$ coupled linearly to the horizon null flux. The dynamical input of Paper I together with its off-diagonal extension in Section~\ref{sec:AAinput} and Appendix~\ref{app:offdiag} establishes this linear readout specifically for the minimally coupled two-scalar Einstein model, where the bulk mutual-information entropy source is already positive (yielding a positive profile $H>0$ that factorizes as a convolution of the gravitational kernel and this source in Section~\ref{sec:qes}); the local positivity derivation of Section~\ref{sec:localpos} thus closes the boundary-to-bulk positivity loop within that model while formulating the exact local test for any broader bulk sector sharing the same eikonal readout. The scalar QES generated by bulk mutual information~\cite{Shape,CW}, with the source normalization matched to the scattering response in Paper I, then provides a concrete application of this dictionary. Its entropy source, gravitational propagation, and probe averaging factorize into a triple convolution, giving exact area and central-thickness benchmarks and sharp deconvolution-free pointwise bounds on the local displacement from a single finite-width scan.

Section~\ref{sec:setup} recalls Paper I's reference geometry, single-insertion preparations, and projected averaged inequality. Section~\ref{sec:inverse} studies single-insertion scale scans, inversion, the causal-shadow area identity, and an explicit analytic counterexample showing why fixed-width averages cannot establish local positivity. Section~\ref{sec:superposition} introduces finite coherent superpositions and off-diagonal fluxes, extends Paper I's $AA$ and cross-remainder bounds to off-diagonal matrix elements, constructs the polynomial derivative probes, and proves moment-matrix completeness for local positivity. Section~\ref{sec:qes} applies this dictionary to the scalar QES, deriving its triple-convolution factorization, exact area and central-thickness benchmarks, and sharp pointwise enclosures. Section~\ref{sec:scope} derives the resulting ordered-limit local positivity constraint, its saturation rigidity, and power-law $G_N$-suppression of both tree-level remainders in a shifted finite boost window, and Section~\ref{sec:discussion} concludes. The appendices present the absolute scalar benchmark and the off-diagonal extension of the $AA$ and cross-channel remainder bounds.

\section{The single-insertion readout from Paper I}
\label{sec:setup}

We collect the reference geometry, single-insertion conventions, and averaged boundary inequality established in Paper I~\cite{CompanionPositivity}. The full-modular construction, independent source normalization, and diagonal complete-graph suppression proofs are derived there. Here we summarize the single-insertion readout before analyzing its resolving power in Section~\ref{sec:inverse} and extending it to coherent superpositions in Section~\ref{sec:superposition}.

\subsection{Reference geometry and single-insertion preparations}
\label{sec:probe}
On the time-symmetric slice of vacuum $\text{CFT}_2$, let $A=(-\infty,0)$, $B=[1-\lambda/2,1+\lambda/2]$ and $R=A\cup B$, with $0<\lambda\ll1$. In units of the AdS radius, the metric is $\dd s^2=z^{-2}(-\dd t^2+\dd x_{\rm P}^2+\dd z^2)$ and $G_N=3/(2c)$, where $c$ is the boundary central charge. The reference extremal curve and causal surface coincide at $x_{\rm P}=0$. The QES branch adjacent to $A$ has displacement
\begin{equation}
 \eta(z)=\GN\lambda^{2\Delta}\eta_*(z)+\cdots,
 \qquad x=\log z,
 \qquad H(x)=e^{-x}\eta_*(e^x)=\frac{\eta_*(z)}z.
 \label{eq:H}
\end{equation}
Positive $\eta$ points towards $B$; $G_N\lambda^{2\Delta}H$ is the leading proper normal width and $\dd x=\dd z/z$ is proper arclength. The reference vacuum and its QES are held fixed while the probe state varies.

We work in the Einstein regime of Paper I with two distinct canonically normalized, minimally coupled real scalars of dimensions $1<\Delta<2$ and $d>2$. The operator $\OO$ is the unique lightest nonidentity primary of any spin. The symmetry $\varphi\mapsto-\varphi$ forbids the cubic vertices $\varphi\phi^2$ and $\varphi^3$. At the order retained here, we set all other direct cubic and quartic matter couplings to zero, including $\varphi^2\phi^2$ and spin-2 contact interactions, and omit additional exchanges and higher-derivative terms. Vacuum correlators use ordered Euclidean-to-Lorentzian continuation in the real-time holographic prescription~\cite{SkenderisVanRees}. The ordered coefficient limits are those specified in Paper I, where both the $AA$ bound and the cross-channel Regge remainder bound are proved for single-insertion preparations at Einstein tree level in the scattering regime of refs.~\cite{CFL,EikonalAdS}.

For the probe primary $\OO_\phi$ of dimension $d$, the normalized single-insertion preparation, its scale-angle coordinates, and its mean CFT energy are
\begin{equation}
 |f_{\tau,\xi}\rangle=(2\tau)^d
 e^{-\tau H_{\rm CFT}}\OO_\phi(0,\xi)|\Omega\rangle,
 \qquad \tau,\xi>0,
 \label{eq:state}
\end{equation}
\begin{equation}
 s=\sqrt{\tau^2+\xi^2},\qquad
 \tau=s\sin\delta,\qquad \xi=s\cos\delta,
 \qquad0<\delta<\frac\pi2,
 \label{eq:prep}
\end{equation}
\begin{equation}
 E_f=\langle f|H_{\rm CFT}|f\rangle=\frac d\tau,
 \label{eq:Ef}
\end{equation}
where $H_{\rm CFT}$ generates boundary time translations and the boundary two-point function is unit-normalized on the Euclidean-preparation sheet, $\langle\OO_\phi(t_1,x_1)\OO_\phi(t_2,x_2)\rangle_\Omega=[(x_1-x_2)^2-(t_1-t_2)^2]^{-d}$.

\subsection{Horizon flux and the projected single-insertion inequality}
\label{sec:dictionary}
Let $u=t+x_{\rm P}$ and $v=t-x_{\rm P}$ be null coordinates, and let $\phi$ be the bulk probe field dual to $\OO_\phi$. The horizon wavefunction $\psi_f(v,z)=\langle\Omega|\phi(u=0,v,z)|f\rangle$ is conveniently written using the positive affine null momentum $p$:
\begin{equation}
 w=\tau+i\xi=ise^{-i\delta},\qquad b=\frac1{\bar w},\qquad
 n_w=\frac{(2\tau b)^d}{\sqrt{2\pi}\Gamma(d)},
 \label{eq:wbn}
\end{equation}
\begin{equation}
 \psi_f(v,z)=\int_0^\infty\dd p\,
 n_w z^d p^{d-1}e^{-p(w+bz^2)}e^{-ipv}.
 \label{eq:momentum}
\end{equation}
Its normalization follows from the canonical bulk scalar action and the unit boundary two-point function, as derived in Paper I. Including the transverse proper measure, the integrated null-flux density is
\begin{equation}
 t_f(z)=\frac1z\int_{\mathbb R}2|\partial_v\psi_f|^2\dd v
 =\frac{E_f}{2}q_d(z;s),\qquad
 q_d(z;s)=\frac{2s^{2d+2}}{B(d,d+1)}
             \frac{z^{2d-1}}{(z^2+s^2)^{2d+1}}.
 \label{eq:flux}
\end{equation}
Here $B(a,b)=\Gamma(a)\Gamma(b)/\Gamma(a+b)$ is Euler's beta function, and $\int_0^\infty q_d(z;s)\dd z=1$. A boost of rapidity $\beta$ multiplies the flux by $e^\beta$.

For any normalized pure global probe state $|F\rangle$ in the common form domain, relative-entropy monotonicity and the full-modular inequality~\cite{CasiniRE,BCFull,ModEnergy,FLPW} give the exact boundary observable used in Paper I,
\begin{equation}
 \PP[F]=(D_R-D_A)+(D_{A^c}-D_{R^c})
 =\langle F|\widehat K_R-\widehat K_A|F\rangle\ge0,
 \label{eq:P}
\end{equation}
where $D_X$ is relative entropy with respect to the vacuum, $K_X=-\log\rho_X^\Omega$, $\widehat K_X=K_X-K_{X^c}$, and we denote the full modular difference operator by $\widehat\PP:=\widehat K_R-\widehat K_A$. Letting $U(\beta)$ denote the unitary vacuum Rindler boost of rapidity $\beta$ and setting $\PP_\beta[F]:=\PP[U(\beta)F]$, the ordered perturbative coefficient and positive Laplace readout are
\begin{align}
 p_F(\beta)&=\lim_{G_N\to0^+}\frac1{G_N}
   \lim_{\lambda\to0^+}\frac{\PP_\beta[F_c]}{\lambda^{2\Delta_c}},
   \label{eq:ordered}\\
 \mathfrak R[F]&=\lim_{\epsilon_A\downarrow0}\epsilon_A
   \int_0^\infty e^{-(1+\epsilon_A)\beta}p_F(\beta)\dd\beta\ge0,
   \label{eq:Abel}
\end{align}
where $|F_c\rangle$ is the corresponding normalized preparation at central charge $c$, $\Delta_c\to\Delta$ is the light dimension, and $\epsilon_A>0$ is the Laplace regulator.

For a single-insertion probe $|f\rangle=|f_{s,\delta}\rangle$, Paper I~\cite[section~5 and appendices~C--D]{CompanionPositivity} decomposed $p_f(\beta)=e^\beta Q^\times[f]+r_\times(\beta)+a_{AA}(\beta)$ into the growing cross response, the finite-boost cross-channel remainder, and the complete same-side $AA$ exchange, proving the uniform exponential decay bounds $|a_{AA}(\beta)|\le C_{AA}e^{-\beta}$ and $e^{-\beta}|r_\times(\beta)|\le C_\times e^{-\beta}$. Both remainders therefore have zero Laplace residue, identifying $\mathfrak R[f]=Q^\times[f]\ge 0$, where the reduced cross coefficient $Q^\times[f]$ is related to the unstripped growing cross response $\PP_\times^{\rm grow}[f]$ by
\begin{equation}
 Q^\times[f]:=\frac{\PP_\times^{\rm grow}[f]}
 {\GN\lambda^{2\Delta}e^\beta},
 \label{eq:Qcross}
\end{equation}
and the independent generalized-entropy source match in Paper I establishes the exact displacement dictionary
\begin{equation}
 \PP_\times^{\rm grow}[f]
       =2\pi e^\beta\int_0^\infty t_f(z)\eta(z)\dd z
       =\pi E_fe^\beta\int_0^\infty q_d(z;s)\eta(z)\dd z.
 \label{eq:dictionary}
\end{equation}
Consequently, Paper I yields the all-scale single-insertion averaged positivity inequality $\int_0^\infty q_d(z;s)\eta_*(z)\dd z\ge 0$ for every $s>0$. In the next section we study what this fixed-width scale scan can and cannot reveal about the local profile $H(x)$, before extending \eqref{eq:Abel} and Paper I's remainder bounds to coherent superpositions in Section~\ref{sec:superposition}.

\section{Scale scans and reconstruction}
\label{sec:inverse}

The cross-channel displacement dictionary \eqref{eq:dictionary} expresses the growing modular response as a radial integral of the coordinate displacement $\eta(z)$ against the probe's horizon flux profile $q_d(z;s)$. To analyze spatial localization, translational invariance, and resolution along the reference surface, we now recast this dictionary into a proper-distance convolution for the dimensionless proper gap profile $H(x)=\eta_*(e^x)/e^x$. In this section, we investigate what can---and cannot---be learned by scanning a single-insertion probe across a continuous range of boundary scales $s=e^y$.

\subsection{From coordinate displacement to a proper-distance convolution}
In terms of the reduced cross coefficient $Q^\times[f]$ defined in \eqref{eq:Qcross}, the displacement dictionary \eqref{eq:dictionary} reads $Q^\times[f]=\pi E_f F_d(s)$, where the coordinate-displacement average $F_d(s)$, the normalization constant $m_d$, and the logarithmic boundary scale $y$ are defined by
\begin{equation}
 F_d(s)=\int_0^\infty q_d(z;s)\eta_*(z)\dd z,\qquad
 m_d=\frac{\Gamma(d+1/2)^2}{\Gamma(d)\Gamma(d+1)},
 \qquad y=\log s.
 \label{eq:Fmd}
\end{equation}
Dividing $F_d(e^y)$ by $m_d e^y$ and using $E_f=d/(e^y\sin\delta)$ defines the shape-normalized proper-displacement readout
\begin{equation}
 G_d(y)=\frac{F_d(e^y)}{m_de^y}
       =\frac{\sin\delta}{\pi d m_d}Q^\times[f_{e^y,\delta}].
 \label{eq:G}
\end{equation}
Notice that multiplying $Q^\times[f_{e^y,\delta}]$ by the known preparation factor $\sin\delta/(\pi d m_d)$ completely removes the dependence on the Euclidean preparation angle $\delta$ and mean energy $E_f$: at leading order, single-insertion scans performed at distinct angles $\delta$ but identical scale $s=e^y$ collapse onto the same universal curve $G_d(y)$. (Throughout, $G_d(y)$ denotes this boundary scan function, not to be confused with the bulk radial Green function $G_L(z,z')$.)

To express $G_d(y)$ directly in terms of the proper profile $H(x)$, we change integration variables in \eqref{eq:Fmd} to the relative proper offset $r=y-x=\log(s/z)$, so that $z=s e^{-r}$, $\dd z=-s e^{-r}\dd r$, and $\eta_*(z)=s e^{-r} H(y-r)$. Reversing the limits of integration and combining the Jacobian factor $e^{-r}$ with the proper conversion factor $e^{-r}$ from $\eta_*(z)/s$ restores reflection symmetry in $r$:
\begin{equation}
 \frac{F_d(s)}s=\frac2{B(d,d+1)}
       \int_{\mathbb R}\frac{H(y-r)}{(2\cosh r)^{2d+1}}\dd r.
 \label{eq:preconvolution}
\end{equation}
Because the unnormalized kernel in \eqref{eq:preconvolution} integrates over $r\in\mathbb{R}$ to $B(d+\frac{1}{2},d+\frac{1}{2})/B(d,d+1)=m_d$, dividing by $m_d$ yields the normalized proper-distance convolution
\begin{equation}
 G_d(y)=(P_d*H)(y),\quad
 P_d(r)=\frac{2}{B(d+\tfrac12,d+\tfrac12)}(2\cosh r)^{-2d-1},
 \quad \int_{\mathbb R}P_d(r)\dd r=1,
 \label{eq:convolution}
\end{equation}
which provides the precise form of \eqref{eq:introtransform}, where throughout this paper $(f*g)(y):=\int_{\mathbb R}f(y-x)g(x)\dd x$ denotes additive convolution on $\mathbb R$. The sampling kernel $P_d(r)$ is strictly positive, even, and centered at $r=0$ ($x=y$), with its width setting the intrinsic spatial resolution of a single-insertion probe of dimension $d$. Varying the boundary scale $y=\log s$ simply translates this sampling window along the reference curve, whereas varying $\delta$ or $\beta$ rescales the overall prefactor prior to normalization without narrowing $P_d(r)$.

\subsection{Ideal uniqueness and unstable inversion}
Because \eqref{eq:convolution} is a translation-invariant convolution on $\mathbb{R}$, its injectivity and conditioning are transparent in Fourier space. Adopting the convention $\widehat H(\omega)=\int_{\mathbb{R}}e^{-i\omega x}H(x)\dd x$ with inverse transform $\frac{1}{2\pi}\int_{\mathbb{R}}e^{i\omega x}\widehat H(\omega)\dd\omega$ and setting $a_d=d+1/2$, an exact beta-function integral (via $t=e^{2r}$) gives the diagonal Fourier multiplier
\begin{equation}
 \widehat G_d(\omega)=T_d(\omega)\widehat H(\omega),\qquad
 T_d(\omega)=\frac{|\Gamma(a_d+i\omega/2)|^2}{\Gamma(a_d)^2}>0.
 \label{eq:transfer}
\end{equation}

Because the Gamma function has no zeros in the complex plane and the argument $a_d+i\omega/2$ ($a_d>5/2$) avoids all poles on the negative real axis, the transfer function $T_d(\omega)$ is strictly positive for all $\omega\in\mathbb{R}$. Consequently, $T_d(\omega)\widehat H(\omega)=0$ a.e.\ implies $\widehat H(\omega)=0$ a.e.: for any fixed probe dimension $d>2$, the forward convolution map $H\mapsto G_d=P_d*H$ is injective on $L^2(\mathbb{R})$, and exact noiseless data $G_d$ in its range uniquely determine $H$ via Fourier division $\widehat H(\omega)=\widehat G_d(\omega)/T_d(\omega)$.

At large frequency $|\omega|\to\infty$, however, Stirling's asymptotic expansion~\cite[section 5.11]{DLMF} yields exponential suppression,
\begin{equation}
 T_d(\omega)\sim\frac{2\pi}{\Gamma(a_d)^2}
          \left(\frac{|\omega|}{2}\right)^{2d}e^{-\pi|\omega|/2}.
 \label{eq:transferasymptotic}
\end{equation}
Unit-normalized profiles $H\in L^2(\mathbb{R})$ whose Fourier transforms are supported at arbitrarily large $|\omega|$ therefore map to readouts $G_d$ of exponentially small $L^2$ norm, rendering the inverse operator $G_d\mapsto H$ unbounded on $L^2(\mathbb{R})$.

\subsection{Finite accuracy and finite aperture}
The exponential damping \eqref{eq:transferasymptotic} implies that reconstructing $H(x)$ from noisy single-insertion data $\widetilde G(y)=G_d(y)+e(y)$ with $L^2$-norm error $\|e\|_2\le\varepsilon$ is severely ill-posed. If we impose a Sobolev smoothness prior $\|H\|_{H^r}\le M$ ($r>0$), where $\|H\|_{H^r}^2:=\frac{1}{2\pi}\int_{\mathbb R}(1+\omega^2)^r|\widehat H(\omega)|^2\dd\omega$, and regularize the inverse via a sharp frequency cutoff $\Lambda>0$,
\begin{equation}
 \widehat H_\Lambda(\omega)
 =\mathbf 1_{|\omega|\le\Lambda}\frac{\widehat{\widetilde G}(\omega)}{T_d(\omega)},
\end{equation}
with $\mathbf 1_{|\omega|\le\Lambda}$ denoting the indicator function of $|\omega|\le\Lambda$, Plancherel's theorem and the monotonic decrease of $T_d(|\omega|)$ yield the $L^2$ reconstruction error bound
\begin{equation}
 \norm{H_\Lambda-H}_2\le
 \frac{\varepsilon}{T_d(\Lambda)}+M(1+\Lambda^2)^{-r/2},
 \label{eq:noisebound}
\end{equation}
where the first term bounds the amplified measurement noise below the cutoff and the second term bounds the truncated high-frequency tail of $H$. Here the strict decrease of $T_d(|\omega|)$ follows immediately from the infinite Weierstrass product representation
\begin{equation}
 T_d(\omega)=\prod_{j=0}^\infty
       \left(1+\frac{\omega^2}{4(a_d+j)^2}\right)^{-1},
 \label{eq:product}
\end{equation}
obtained by iterating the Gamma recurrence $\Gamma(z+1)=z\Gamma(z)$ and noting that $T_{d+n}(\omega)\to 1$ as $n\to\infty$ at fixed $\omega$. Balancing the two terms in \eqref{eq:noisebound} selects a cutoff $\Lambda=(2/\pi)[\log(M/\varepsilon)+(2d-r)\log\log(M/\varepsilon)+O(1)]$ and yields only a logarithmic convergence rate $\|H_\Lambda-H\|_2=O(M[\log(M/\varepsilon)]^{-r})$ as $\varepsilon\to 0$.

Similarly, while the exponential decay of $T_d(\omega)$ guarantees that $G_d(y)$ extends to a holomorphic function in the strip $|\operatorname{Im}y|<\pi/2$---so that exact data on any open interval of $y$ uniquely determine $G_d$ on all of $\mathbb{R}$ by analytic continuation---such analytic continuation is exponentially unstable against noise, and any finite discrete set of scale samples leaves an infinite-dimensional nullspace in $L^2(\mathbb{R})$.

\subsection{Moment identities and the spatial area of the causal shadow}
By contrast with pointwise deconvolution, global low-order moments of $H(x)$ can be extracted directly from the single-insertion scan $G_d(y)$ without any inverse filtering. Whenever the relevant absolute moments are finite and $M_0:=\int_{\mathbb{R}}H(x)\dd x\ne 0$, the unit-normalized even kernel $P_d$ preserves the total integral and centroid while shifting the variance additively:
\begin{align}
 \int G_d\dd y&=\int H\dd x=M_0,\qquad \mu(G_d)=\mu(H),\\
 \Var(G_d)&=\Var(H)+\tfrac12\psi_1(d+\tfrac12).
 \label{eq:moments}
\end{align}
Here $\mu(F)=(\int xF(x)\dd x)/(\int F)$ and $\Var(F)=(\int (x-\mu(F))^2F(x)\dd x)/(\int F)$. For sign-changing profiles these are signed moment identities, with no claim that the variance is nonnegative. The function $\psi_1(z)=\frac{\dd^2}{\dd z^2}\log\Gamma(z)$ is the trigamma function and $\sigma_d^2=\frac{1}{2}\psi_1(d+\frac{1}{2})\sim 1/(2d)$ is the variance of $P_d(r)$.

Remarkably, the zeroth moment $M_0$ admits a direct bulk geometric interpretation. On the time-symmetric slice $t=0$ of Poincar\'e $\text{AdS}_3$, the induced two-dimensional spatial area element is $\dd x_{\rm P}\,\dd z/z^2$. Consequently, the oriented strip integral $\mathcal A_{\rm strip}$ between the reference causal horizon $x_{\rm P}=0$ and the displaced QES curve $x_{\rm P}=\eta(z)$ is given at leading order by
\begin{equation}
 \begin{aligned}
 \mathcal A_{\rm strip}
 &=\int_0^\infty\frac{\eta(z)}{z^2}\dd z\\
 &=\GN\lambda^{2\Delta}\int_{\mathbb R}H(x)\dd x
  =\GN\lambda^{2\Delta}\int_{\mathbb R}G_d(y)\dd y
    +o(\GN\lambda^{2\Delta}).
 \end{aligned}
 \label{eq:area}
\end{equation}
For $H\ge0$, this oriented integral is the leading spatial area of the causal-shadow strip on the $t=0$ slice. Before its sign is established, the same formula gives an oriented area, not the unsigned area of a sign-changing strip. In either case the zeroth moment requires no deconvolution.

\subsection{Why single-insertion positivity is not local positivity}
For $H(x)\ge0$, integrating the nonnegative function $P_d(y-x)H(x)$ gives $G_d(y)=(P_d*H)(y)\ge 0$ for all $y\in\mathbb{R}$. However, the converse is false: because $P_d(r)$ has a fixed nonzero width of order $d^{-1/2}$, a localized negative dip in $H(x)$ can be completely overwhelmed by surrounding positive support at every scan center $y$. To see this explicitly, consider the smooth, localized sign-changing test profile
\begin{equation}
 H_{\rm test}(x)=\sech x-2\chi(x/0.1),\qquad
 \chi(u)=\begin{cases}
 e^{1-1/(1-u^2)},&|u|<1,\\0,&|u|\ge1,
 \end{cases}
 \label{eq:notch}
\end{equation}
which has a negative central notch $H_{\rm test}(0)=-1$ supported on $|x|<0.1$. For a scalar probe of dimension $d=3$, convexity of $u\mapsto u^{-(2d+1)}$ applied to $\tfrac12[\cosh(y-x)+\cosh(y+x)]=\cosh y\cosh x$ gives the symmetrized lower bound
\begin{equation}
 \frac{P_d(y-x)+P_d(y+x)}{2P_d(y)}\ge\sech^{2d+1}x.
 \label{eq:convexitybound}
\end{equation}
Using the evenness of $\sech x$ for the positive term, together with $|(\log P_3)'(r)|=7|\tanh r|\le 7$ (so $P_3(y-x)/P_3(y)\le e^{0.7}$ on $|x|<0.1$) and $0\le\chi\le 1$ for the negative notch, yields at every scan center $y\in\mathbb{R}$
\begin{equation}
 \frac{(P_3*H_{\rm test})(y)}{P_3(y)}
 \ge\int_{\mathbb R}\sech^8x\,\dd x-0.4e^{0.7}
 =\frac{32}{35}-0.4e^{0.7}>0.10878.
 \label{eq:single-insertionpositive}
\end{equation}
Consequently, the single-insertion readout $(P_3*H_{\rm test})(y)$ is strictly positive for \emph{all} $y\in\mathbb{R}$, completely concealing the negative notch at $x=0$.

At a purely mathematical level, one can narrow the kernel $P_d$ to $P_{d+n}$ by applying a $2n$-th order differential operator in the scan variable $y$. Iterating the Fourier product \eqref{eq:product} shows that
\begin{equation}
 \mathcal D_{d,n}G_d=P_{d+n}*H,\qquad
 \mathcal D_{d,n}=\prod_{j=0}^{n-1}
       \left(1-\frac{\partial_y^2}{4(d+1/2+j)^2}\right).
 \label{eq:diffhierarchy}
\end{equation}
Because $\{P_{d+n}\}_{n\ge 0}$ forms a positive approximate identity on $\mathbb{R}$ with variance $\frac{1}{2}\psi_1(d+n+\frac{1}{2})=O(n^{-1})\to 0$, uniform continuity implies that for any $H\in C_0(\mathbb{R})\cap L^1(\mathbb{R})\cap L^2(\mathbb{R})$ (where $C_0(\mathbb{R})$ denotes continuous functions vanishing as $|x|\to\infty$),
\begin{equation}
 H\ge0\quad\Longleftrightarrow\quad
 \mathcal D_{d,n}G_d\ge0\ \text{for all }n,y.
 \label{eq:diffcriterion}
\end{equation}
Similarly, if $(P_{d+n}*H)(y)\ge -\varepsilon_n$ and supremum norm $\|H''\|_\infty\le M_2$, Taylor expanding $H$ to second order gives the finite-order lower bound
\begin{equation}
 H(y)\ge-\varepsilon_n-\frac{M_2}{4}\psi_1(d+n+\tfrac12).
 \label{eq:finitecurvature}
\end{equation}
Crucially, however, \eqref{eq:diffhierarchy}--\eqref{eq:diffcriterion} do \emph{not} by themselves give a positivity criterion implemented by physical boundary states: the CFT spectrum need not contain scalar primaries of dimension $d+n$ for every integer $n\ge 1$, and classical post-processing with $\partial_y^2$ takes signed differences of nonnegative single-insertion measurements $G_d(y)\ge 0$. Consequently, the scan constraint $G_d(y)\ge 0$ alone does \emph{not} imply
\[
 \left(1-\frac{\partial_y^2}{(2d+1)^2}\right)G_d(y)\ge 0,
\]
and numerical differentiation also amplifies high-frequency noise (at fixed order $n$, the Fourier multiplier grows as $|\omega|^{2n}$, whereas exponential ill-conditioning refers to the full inverse transform). To formulate local positivity criteria within the boundary Hilbert space using a \emph{single} scalar species $\OO_\phi$, we must differentiate the \emph{state} rather than the \emph{data}: forming a derivative state $|\partial f\rangle=\lim_{h\to 0}(|f_{y+h}\rangle-|f_y\rangle)/h$ inside the Hilbert space first and only then evaluating $\widehat\PP$ preserves exact positivity,
\[
 \langle\partial f|\widehat\PP|\partial f\rangle\ge 0
 \quad\Longrightarrow\quad
 Q^\times[\partial f/\norm{\partial f}]\ge 0,
\]
using destructive quantum interference between coherent superpositions of the same scalar species $\OO_\phi$ to narrow the horizon null-flux sampling profile arbitrarily, as we now construct in Section~\ref{sec:superposition}.

\section{Coherent superpositions and local positivity}
\label{sec:superposition}

The central goal of this section is to upgrade Paper I's fixed-width single-insertion inequality $G_d(y)\ge 0$ to local nonnegativity $H(x)\ge 0$ using only the single scalar species $\OO_\phi$. Whereas single-insertion probes merely translate the fixed positive kernel $P_d(y-x)$, superposing distinct Euclidean preparations introduces off-diagonal interference in the horizon null-energy flux while staying strictly inside the physical boundary Hilbert space. We carry out the proof in three direct steps: we first derive the off-diagonal horizon flux and extend Paper I's complete-graph $AA$ and cross-channel remainder suppression bounds to finite coherent superpositions (Appendix~\ref{app:offdiag}), transferring exact boundary positivity to the geometric quadratic form $Q^\times[F/\norm F]\ge 0$; we next show that finite derivatives with respect to $(\tau,\xi)$ act as commuting polynomial multipliers along the compactified horizon, producing normalized states whose physical readouts have the exact same sign as arbitrary polynomial-weighted averages $\int_{-1}^1 H|p|^2\dd\mu_d$; and we then convert these weights into a moment-matrix hierarchy whose completeness establishes exact equivalence to $H\ge 0$ almost everywhere at a single scan center.

\subsection{Coherent superpositions, off-diagonal flux, and two norms}
Consider a finite superposition of single-insertion Euclidean states prepared at distinct parameters $(\tau_i,\xi_i)$ ($i=1,\dots,N_f$), writing $|f_i\rangle:=|f_{\tau_i,\xi_i}\rangle$ and $|F\rangle=\sum_i c_i|f_i\rangle$. Using the unit-normalized boundary two-point function with time separation $-i(\tau_i+\tau_j)$ and spatial separation $\xi_i-\xi_j$, multiplying the unnormalized overlap $\langle\widetilde f_i|\widetilde f_j\rangle=[(\tau_i+\tau_j)^2+(\xi_i-\xi_j)^2]^{-d}$ by $(2\tau_i)^d(2\tau_j)^d$ gives the physical boundary Gram matrix
\begin{equation}
 S_{ij}:=\langle f_i|f_j\rangle
 =\left[\frac{4\tau_i\tau_j}
 { (\tau_i+\tau_j)^2+(\xi_i-\xi_j)^2}\right]^d,
 \label{eq:S}
\end{equation}
so that $S_{ii}=1$ and $\norm{F}^2=c^\dagger S c$. By linearity of the one-particle bulk field matrix element, the horizon wavefunction of $|F\rangle$ is $\psi_F(v,z)=\sum_i c_i\psi_i(v,z)$, and its normalized horizon null-flux density becomes a Hermitian quadratic form:
\begin{equation}
 t_{F/\norm F}(z)=\frac{\sum_{ij}c_i^*c_jt_{ij}(z)}{c^\dagger S c},
 \qquad
 t_{ij}(z)=\frac1z\int_{\mathbb R}
 2\partial_v\psi_i^*\partial_v\psi_j\dd v.
 \label{eq:interferenceflux}
\end{equation}
Whereas the diagonal entries $t_{ii}(z)$ reproduce the single-insertion profiles \eqref{eq:flux}, the off-diagonal interference kernels $t_{ij}(z)$ ($i\ne j$) depend on the relative preparation phases and reshape the radial sampling weight. Introducing the complex pairwise preparation combinations
\begin{equation}
 \mathfrak a_{ij}=\bar w_i+w_j,
 \qquad \mathfrak b_{ij}=\bar b_i+b_j,
 \label{eq:abij}
\end{equation}
and inserting the positive null-momentum representation \eqref{eq:momentum} into \eqref{eq:interferenceflux} (with $p$ and $q$ denoting the null momenta of $\psi_j$ and $\psi_i^*$), the longitudinal $v$-integral yields $2\pi\delta(p-q)$ and the remaining $p$-integral evaluates to
\begin{equation}
 t_{ij}(z)=4\pi n_i^*n_j\Gamma(2d+1)
       \frac{z^{2d-1}}{(\mathfrak a_{ij}+\mathfrak b_{ij}z^2)^{2d+1}},
 \qquad n_i=n_{w_i}.
 \label{eq:offdiagflux}
\end{equation}
By construction, $t_{ij}(z)$ is a positive semidefinite Gram kernel at each $z>0$ ($t_{ji}=t_{ij}^*$, $|t_{ij}|^2\le t_{ii}t_{jj}$), setting $i=j$ recovers \eqref{eq:flux}, and its smooth preparation dependence allows finite differences to converge to the derivative probes below.

Retaining the momentum integral before integrating over $p$ and passing to proper arclength $x=\log z$ gives the factorized representation $e^{2x}t_{ij}(e^x)=\int_0^\infty W_i(p,x)^*W_j(p,x)\dd p$, in terms of the horizon mode functions
\begin{equation}
 W_i(p,x)=\sqrt{4\pi}\,n_i p^d e^{(d+1/2)x}
               e^{-p(w_i+b_i e^{2x})}.
 \label{eq:W}
\end{equation}
Because the gravitational shockwave shift $h(z)=2\pi\eta(z)$ is sourced solely by the fixed light-scalar pair and couples linearly to the probe's horizon stress tensor $T_{vv}^\phi$ (verified directly from the off-diagonal Mellin discontinuity in Appendix~\ref{app:offdiag}), pairing $t_{ij}(z)$ with the stripped QES displacement $\eta_*(z)=e^x H(x)$ gives the entries $(Q_\times)_{ij}=Q^\times_{ij}$ of the Hermitian cross-response matrix $Q_\times$,
\begin{equation}
 Q^\times_{ij}=2\pi\int\dd x\dd p\,H(x)W_i(p,x)^*W_j(p,x),
 \label{eq:crossform}
\end{equation}
and the reduced cross-channel readout of any normalized superposition $|F\rangle/\norm F$,
\begin{equation}
 Q^\times[F/\norm F]
 =\frac{c^\dagger Q_\times c}{c^\dagger Sc}
 =\frac{2\pi}{c^\dagger Sc}
   \int\dd x\dd p\,H(x)\left|\sum_i c_iW_i(p,x)\right|^2,
 \label{eq:coherentform}
\end{equation}
where $x\in\mathbb R$ and $p>0$. Thus the same profile $H$ is paired with a sampling weight $|\sum_i c_i W_i|^2$ that can be shaped by quantum interference. Setting $H=1$ defines the horizon response inner product,
\begin{equation}
 \langle W_i,W_j\rangle
 =\frac{2\pi n_i^*n_j\Gamma(d+1/2)^2}
 {(\mathfrak a_{ij}\mathfrak b_{ij})^{d+1/2}},
 \qquad \norm{W_i}^2=\frac{dm_d}{2\sin\delta_i},
 \label{eq:responseGram}
\end{equation}
which differs from the physical boundary overlap $S_{ij}$ in \eqref{eq:S}. Their distinction is crucial: shaping the numerator of \eqref{eq:coherentform} also changes the physical boundary normalization of the state. Before carrying both inner products through the derivative construction, we establish that exact boundary positivity transfers to $Q^\times[F/\norm F]\ge 0$ on every fixed finite superposition space.

\subsection{Off-diagonal \texorpdfstring{$AA$}{AA} suppression, remainder bounds, and projection}
\label{sec:AAinput}

To transfer the exact boundary inequality $\PP[F]\ge 0$ in \eqref{eq:P} to the coherent quadratic form \eqref{eq:coherentform}, we must extend Paper I's single-insertion $AA$ and cross-channel remainder bounds to off-diagonal matrix elements. On a fixed $N_f$-dimensional subspace of linearly independent single-insertion states $\{|f_i\rangle\}_{i=1}^{N_f}$ with positive definite Gram matrix $S\succ 0$ (smallest eigenvalue $s_{\min}>0$; throughout, $A\succ 0$ and $A\succeq 0$ denote positive definite and positive semidefinite Hermitian matrices), polarizing the ordered perturbative coefficient \eqref{eq:ordered} defines a Hermitian matrix $\mathsf p(\beta)$ with $p_F(\beta)=c^\dagger\mathsf p(\beta)c/(c^\dagger S c)\ge 0$, or equivalently $S^{-1/2}\mathsf p(\beta)S^{-1/2}\succeq 0$. As in Paper I~\cite{CompanionPositivity}, microcausality eliminates $K_B$ and $\delta K_{BB}$, leaving two surviving channels at Einstein tree order:
\begin{equation}
 \mathsf p(\beta)=\mathsf p_\times(\beta)+\mathsf a_{AA}(\beta),
 \qquad
 \mathsf p_\times(\beta)=e^\beta Q_\times+\mathsf r_\times(\beta),
 \label{eq:AAdecomposition}
\end{equation}
where $\mathsf p_\times$ is the complete cross-response matrix, $\mathsf a_{AA}$ is the complete same-side Einstein contribution with both light insertions in $A$ (with $\GN\lambda^{2\Delta}$ stripped from both), $Q_\times$ is the cross-response matrix \eqref{eq:crossform}, and $\mathsf r_\times:=\mathsf p_\times-e^\beta Q_\times$ is the finite-boost cross-channel remainder at the same tree order.

In Appendix~\ref{app:offdiag}, we extend Paper I's complete tree-level graviton exchange estimates~\cite[section~5.2 and appendices~C--D]{CompanionPositivity}---which retain all regular and logarithmic scalar double-trace conformal families~\cite{WittenBlocks} on the carried Lorentzian sheets~\cite{SkenderisVanRees}---from diagonal preparations ($s_i=s_j,\ \delta_i=\delta_j$) to general off-diagonal pairs $(s_i,\delta_i)\ne(s_j,\delta_j)$. For the same-side $AA$ channel, transferring the boost $\beta$ to the exact Hermitian modular kernel and keeping both ordered words on the Wightman sheet through the mixed corners and one-leg edges establishes the same uniform graph envelope $|F_{\rm word}(a,g)|\le C_{ij}G_N\min(L^2,1)$ across all $N_f^2$ pairs (where $a,g$ are the boost-shifted $AA$ modular coordinates, $L=|a|+|g|$, and $F_{\rm word}$ is the normalized connected four-point function for each commutator ordering, as in Paper I~\cite[section~5.2]{CompanionPositivity}), yielding the entrywise bound
\begin{align}
 |\mathsf a_{AA,ij}(\beta)|
 &\le C_{ij}e^{-\beta}B(2\Delta,2\Delta)
       \int_0^\infty\frac{\min(L^2,1)}{L^2}\dd L\nonumber\\
 &=2C_{ij}B(2\Delta,2\Delta)e^{-\beta}.
 \label{eq:AAentry}
\end{align}
In terms of the matrix operator (spectral) norm $\|\cdot\|_{\rm op}$, the finite-matrix norm inequality gives
\begin{equation}
 \big\|S^{-1/2}\mathsf a_{AA}(\beta)S^{-1/2}\big\|_{\rm op}
 \le \frac{N_f}{s_{\min}}\max_{ij}|\mathsf a_{AA,ij}(\beta)|
 \le C_{AA} e^{-\beta},
 \qquad \beta\ge 0.
 \label{eq:AAprobeBound}
\end{equation}
Writing $v=S^{1/2}c$ for $|F\rangle=\sum_i c_i|f_i\rangle$, its normalized $AA$ response therefore obeys
\begin{equation}
 \left|\frac{c^\dagger\mathsf a_{AA}(\beta)c}{c^\dagger Sc}\right|
 =\left|\frac{v^\dagger S^{-1/2}\mathsf a_{AA}(\beta)S^{-1/2}v}{v^\dagger v}\right|
 \le C_{AA} e^{-\beta},
 \label{eq:AAsuperposition}
\end{equation}
controlling interference between distinct preparations uniformly in $\beta\ge 0$. Similarly, Appendix~\ref{app:offdiag} extends Paper I's exact Mellin discontinuity and dominated-convergence proof for the cross-channel remainder to general off-diagonal pairs (accounting for the off-diagonal phase $\arg(\mathfrak a_{ij}\mathfrak b_{ij})=\delta_i-\delta_j\in(-\pi/2,\pi/2)$), proving that $|\mathsf r_{\times,ij}(\beta)|\le C_{\times,ij}$ is locally integrable and uniformly bounded on $[0,\infty)$ and hence satisfies the physical operator-norm bound
\begin{equation}
 h_\times(\beta):=e^{-\beta}
 \big\|S^{-1/2}\mathsf r_\times(\beta)S^{-1/2}\big\|_{\rm op}
 \le C_\times e^{-\beta},\qquad \beta\ge 0.
 \label{eq:crossRemainderBoundII}
\end{equation}

Applying the positive Laplace projection \eqref{eq:Abel} to the Hermitian response matrix $S^{-1/2}\mathsf p(\beta)S^{-1/2}\succeq 0$ immediately projects both remainders to zero as $\epsilon_A\downarrow 0$:
\begin{align}
 \left\|\epsilon_A\int_0^\infty e^{-(1+\epsilon_A)\beta}
 S^{-1/2}\mathsf a_{AA}(\beta)S^{-1/2}\dd\beta\right\|_{\rm op}
 &\le C_{AA}\frac{\epsilon_A}{2+\epsilon_A}\longrightarrow 0,
 \label{eq:AAAbelZero}\\
 \left\|\epsilon_A\int_0^\infty e^{-(1+\epsilon_A)\beta}
 S^{-1/2}\mathsf r_\times(\beta)S^{-1/2}\dd\beta\right\|_{\rm op}
 &\le\epsilon_A\int_0^\infty e^{-\epsilon_A\beta}h_\times(\beta)\dd\beta\nonumber\\
 &\le C_\times\frac{\epsilon_A}{1+\epsilon_A}\longrightarrow 0.
 \label{eq:crossRemainderAbel}
\end{align}
Because the cone of positive semidefinite matrices on a finite-dimensional space is closed, taking $\epsilon_A\downarrow 0$ yields $\mathfrak R=S^{-1/2}Q_\times S^{-1/2}\succeq 0$, i.e.\
\begin{equation}
 \mathfrak R[F/\norm F]=Q^\times[F/\norm F]
 =\frac{2\pi}{c^\dagger Sc}
   \int_0^\infty\!\dd p\int_{\mathbb R}\!\dd x\,
   H(x)\left|\sum_i c_iW_i(p,x)\right|^2\ge 0
 \label{eq:AbelSuperposition}
\end{equation}
for every fixed finite superposition $|F\rangle=\sum_i c_i|f_i\rangle$. Moreover, as shown in Appendix~\ref{app:offdiag}, the integral majorants hold on small polydisks in the independent bra and ket preparation parameters. Cauchy estimates then preserve the $O(e^{-\beta})$ decay directly under every fixed finite order of preparation derivatives (with constants depending on that order and on the derivative-space Gram matrix). Independently, the derivation in Section~\ref{sec:localpos} operates purely on fixed finite-dimensional spaces of single-insertion preparations first---taking the coefficient and Laplace limits at fixed finite-difference step $h\ne 0$ and sending $h\to 0$ only inside the smooth cross-channel readout \eqref{eq:offdiagflux}---so our main local positivity result does not even rely on differentiating the unprojected bounds.

\subsection{Preparation derivatives as polynomial multipliers}
Section~\ref{sec:AAinput} established $Q^\times[F/\norm F]\ge 0$ for every finite superposition of single-insertion $\OO_\phi$ preparations. To turn this freedom into a proof of local positivity $H(x)\ge 0$, we need a systematic basis of superpositions that replaces the fixed kernel $P_d(y-x)$ with arbitrary nonnegative polynomial weights $|p(t)|^2$ along the horizon. We construct these derivative states here, evaluate their normalized readouts in the next subsection, and prove moment-matrix completeness for $H\ge 0$ in Section~\ref{sec:localtests}.

At a single reference preparation $w=\tau+i\xi$, we consider infinitesimal superpositions---finite-order derivatives with respect to $(\tau,\xi)$---of the stripped single-insertion state
\begin{equation}
 |g_w\rangle=|f_w\rangle/n_w
 =\sqrt{2\pi}\Gamma(d)\bar w^d e^{-\tau H_{\rm CFT}}
                \OO_\phi(0,\xi)|\Omega\rangle.
 \label{eq:g}
\end{equation}
Dividing out $n_w$ removes the $w$-dependent normalization prefactor from the horizon mode \eqref{eq:W} (recall $b=1/\bar w$ from \eqref{eq:wbn}), isolating pure double-Laplace exponential dependence $W_w(p,x)/n_w=\sqrt{4\pi}\,p^d e^{(d+1/2)x}e^{-wU-V/\bar w}$ in the independent positive variables
\begin{equation}
 U=p,\qquad V=pz^2=pe^{2x},
 \label{eq:UV}
\end{equation}
where $U=p$ is the horizon null momentum and the ratio $V/U=z^2=e^{2x}$ isolates the bulk proper position $x=\log z$. Consequently, the commuting complex differential operators $-\partial_w$ and $D_V:=\bar w^2\partial_{\bar w}$ act on the horizon wavefunction algebraically as multiplication by $U$ and $V$:
\begin{equation}
 -\partial_w\longleftrightarrow U,
 \qquad D_V:=\bar w^2\partial_{\bar w}\longleftrightarrow V,
 \quad
 \partial_w=\tfrac12(\partial_\tau-i\partial_\xi),\quad
 \partial_{\bar w}=\tfrac12(\partial_\tau+i\partial_\xi).
 \label{eq:multipliers}
\end{equation}
Because $\tau>0$, every fixed finite-order derivative of $|g_w\rangle$ is a smooth, normalizable vector in the boundary Hilbert space, approximated in both boundary norm and the cross-channel quadratic form \eqref{eq:coherentform} by finite-difference superpositions of nearby Euclidean preparations, while the bulk scalar mass $m_\phi^2=d(d-2)$ remains unchanged.

To separate the null-momentum scale from the spatial position along the horizon at a chosen reference preparation $(s,\delta)$ (with $y=\log s$, $\tau=s\sin\delta$, and $|w|^2=s^2$), observe that the squared exponential damping factor in the horizon density is simply $|e^{-wU-V/\bar w}|^2=e^{-2\tau U-(2\tau/s^2)V}$, while the ratio $(2\tau/s^2)V/(2\tau U)=e^{2(x-y)}$ depends solely on the proper-distance offset $x-y$. We therefore rescale $(U,V)$ to dimensionless variables $(\mathsf u,\mathsf v)$ and pass to the total radial damping variable $\mathsf q$ and the compactified proper-distance coordinate $t$:
\begin{equation}
 \mathsf u=2\tau U,\quad \mathsf v=\frac{2\tau}{s^2}V,
 \quad\mathsf q=\mathsf u+\mathsf v,\quad
 t=\frac{\mathsf v-\mathsf u}{\mathsf q}=\tanh(x-y).
 \label{eq:polycoords}
\end{equation}
In these variables, $|e^{-wU-V/\bar w}|^2=e^{-\mathsf q}$ depends only on $\mathsf q>0$, and $t=\tanh(x-y)\in(-1,1)$ maps the entire reference curve $x\in(-\infty,\infty)$ diffeomorphically onto $(-1,1)$, centered at $x=y$ ($t=0$). (Throughout Section~\ref{sec:superposition}, $t=\tanh(x-y)$ denotes this compactified spatial coordinate along the horizon, distinct from the boundary time coordinate of Section~\ref{sec:setup}.)

Because $t=(\mathsf v-\mathsf u)/(\mathsf u+\mathsf v)$ is a rational function of $(\mathsf u,\mathsf v)$, multiplying any polynomial $p(t)$ of degree at most $n$ by $\mathsf q^n=(\mathsf u+\mathsf v)^n$ clears all denominators and produces its homogeneous lift
\begin{equation}
 \Pi_{n,p}(\mathsf u,\mathsf v)
   =(\mathsf u+\mathsf v)^n
       p\!\left(\frac{\mathsf v-\mathsf u}{\mathsf u+\mathsf v}\right),
 \label{eq:polylift}
\end{equation}
which is an ordinary homogeneous polynomial of degree $n$ in $(\mathsf u,\mathsf v)$. Lifting all polynomials of degree at most $n$ to the \emph{same} homogeneous degree $n$ also guarantees that every state in the resulting $(n+1)$-dimensional subspace carries the identical radial factor $\mathsf q^n$, allowing the radial $\mathsf q$-integral to factor out uniformly across the moment matrices in Section~\ref{sec:localtests}. Substituting $\mathsf u\mapsto 2\tau(-\partial_w)$ and $\mathsf v\mapsto (2\tau/s^2)D_V$ in $\Pi_{n,p}(\mathsf u,\mathsf v)$ and acting on $|g_w\rangle$ defines a physical boundary derivative state $|g_{n,p;y,\delta}\rangle$ whose horizon wavefunction is multiplied by $\mathsf q^n p(t)$. In this substitution, the numerical reference factors $2\tau$ and $2\tau/s^2$ are held constant during differentiation and evaluated at the reference preparation, whereas $\bar w^2$ stays inside each power of the vector field $D_V=\bar w^2\partial_{\bar w}$ (so that $D_V^k=(\bar w^2\partial_{\bar w})^k\ne\bar w^{2k}\partial_{\bar w}^k$). Each derivative state $|g_{n,p;y,\delta}\rangle$ is realized as the $h\to 0$ limit of iterated symmetric finite differences of allowed Euclidean preparations (with $h$ chosen small enough that every stencil point lies in $\tau,\xi>0$, e.g.\ $n|h|<\min(\tau,\xi)$ for an $n$-th order iterated difference), built from
\begin{equation}
 \partial_\tau|g_{\tau,\xi}\rangle
 =\lim_{h\to0}\frac{|g_{\tau+h,\xi}\rangle-|g_{\tau-h,\xi}\rangle}{2h}
 \label{eq:preparationdifference}
\end{equation}
and analogously for $\partial_\xi|g_{\tau,\xi}\rangle$; hence every $|g_{n,p;y,\delta}\rangle$ is a limit of fixed finite superpositions governed by Section~\ref{sec:AAinput}.

\subsection{Two-measure normalization and the polynomial readout}
As noted in \eqref{eq:S} and \eqref{eq:responseGram}, modifying the probe state changes both the horizon response (numerator) and the physical boundary norm (denominator), which are controlled by distinct inner products. To evaluate both simultaneously for $|g_{n,p;y,\delta}\rangle$, we express the stripped boundary overlap $\langle g_i|g_j\rangle$ and the unnormalized stripped cross response $R^\times_{ij}:=Q^\times_{ij}/(n_i^*n_j)$ (so that $Q^\times[g/\norm g]=R^\times[g]/\langle g|g\rangle$) in the common double-Laplace variables $(U,V)$. Using $\dd U\,\dd V=2pe^{2x}\dd p\,\dd x$ and $x=\frac{1}{2}\log(V/U)$ in \eqref{eq:W}--\eqref{eq:crossform}, the two kernels share the identical exponential factor $e^{-\mathfrak a_{ij}U-\mathfrak b_{ij}V}$ (with $\mathfrak a_{ij},\mathfrak b_{ij}$ from \eqref{eq:abij}) and differ only by a factor of $(UV)^{1/2}$:
\begin{align}
 \langle g_i|g_j\rangle
 &=\frac{2\pi\Gamma(d)^2}{(\mathfrak a_{ij}\mathfrak b_{ij})^d}
 =2\pi\int_0^\infty\!\dd U\dd V\,(UV)^{d-1}
                      e^{-\mathfrak a_{ij}U-\mathfrak b_{ij}V},\nonumber\\
 R^\times_{ij}
 &=4\pi^2\int_0^\infty\!\dd U\dd V\,(UV)^{d-1/2}
   H\!\left(\tfrac12\log\tfrac{V}{U}\right)
   e^{-\mathfrak a_{ij}U-\mathfrak b_{ij}V}.
 \label{eq:laplaceGram}
\end{align}
At the common reference preparation $i=j=w$ (where $\mathfrak a_{ww}U+\mathfrak b_{ww}V=2\tau U+\frac{2\tau}{s^2}V=\mathsf u+\mathsf v=\mathsf q$), changing variables to $(\mathsf q,t)$ via $\mathsf u=\frac{1}{2}\mathsf q(1-t)$, $\mathsf v=\frac{1}{2}\mathsf q(1+t)$, and $\dd\mathsf u\,\dd\mathsf v=\frac{1}{2}\mathsf q\,\dd\mathsf q\,\dd t$ gives
\begin{equation}
 UV=\frac{s^2}{4\tau^2}\mathsf u\mathsf v
   =\frac{s^2}{16\tau^2}\mathsf q^2(1-t^2),
 \qquad
 \frac12\log\frac{V}{U}=x=y+\arctanh t.
 \label{eq:UVtoqt}
\end{equation}
Accordingly, $(UV)^{d-1}\dd U\,\dd V\propto\mathsf q^{2d-1}(1-t^2)^{d-1}\dd\mathsf q\,\dd t$ in the boundary overlap, and multiplying by $(UV)^{1/2}\propto\mathsf q(1-t^2)^{1/2}$ gives $\mathsf q^{2d}(1-t^2)^{d-1/2}\dd\mathsf q\,\dd t$ in the horizon response. The radial $\mathsf q$-integrals factor out immediately as $\Gamma(2d)$ and $\Gamma(2d+1)$, leaving two distinct unit-normalized Jacobi probability measures on $t\in(-1,1)$:
\begin{equation}
 \dd\mu_d(t)=\frac{(1-t^2)^{d-1/2}}{B(1/2,d+1/2)}\dd t,
 \qquad
 \dd\nu_d(t)=\frac{(1-t^2)^{d-1}}{B(1/2,d)}\dd t.
 \label{eq:measures}
\end{equation}
Because $t=\tanh(x-y)$ satisfies $1-t^2=\sech^2(x-y)$ and $\dd t=\sech^2(x-y)\,\dd x$, the horizon response measure $\dd\mu_d(t)$ is identically the single-insertion kernel of \eqref{eq:convolution}:
\begin{equation}
 \dd\mu_d(t)=P_d(x-y)\,\dd x,
 \qquad t=\tanh(x-y)\in(-1,1).
 \label{eq:muPd}
\end{equation}
The two measures play complementary physical roles:
\begin{center}
\small
\begin{tabular}{@{}lll@{}}\toprule
Measure & Origin & What it normalizes\\\midrule
$\dd\nu_d$ & boundary Hilbert-space inner product & the physical preparation state\\
$\dd\mu_d$ & horizon response inner product & the shape-sampling profile\\
\bottomrule\end{tabular}
\end{center}

Let $\mathcal S_{n,p}:=\langle g_{n,p;y,\delta}|g_{n,p;y,\delta}\rangle$ denote the boundary norm (with $\mathcal S_{0,1}=\langle g_w|g_w\rangle$), and define the normalized readout $Q_n^\times[p;y]:=Q^\times[g_{n,p;y,\delta}/\sqrt{\mathcal S_{n,p}}]$. Multiplying the horizon wavefunction by $\Pi_{n,p}(\mathsf u,\mathsf v)=\mathsf q^n p(t)$ inserts $|\Pi_{n,p}|^2=\mathsf q^{2n}|p(t)|^2$ into both integrals of \eqref{eq:laplaceGram}:
\begin{itemize}
 \item in the boundary overlap, it shifts the radial integral from $\Gamma(2d)$ to $\Gamma(2d+2n)$ and weights $|p(t)|^2$ by $\dd\nu_d(t)$;
 \item in the horizon response $R^\times$, it shifts the radial integral from $\Gamma(2d+1)$ to $\Gamma(2d+2n+1)$ and pairs $H(y+\arctanh t)|p(t)|^2$ against $\dd\mu_d(t)$.
\end{itemize}
Dividing the horizon response by the boundary norm multiplies the $n=0$ prefactor $\pi d m_d/\sin\delta$ of \eqref{eq:G} by the radial Gamma ratio $\frac{\Gamma(2d+2n+1)/\Gamma(2d+1)}{\Gamma(2d+2n)/\Gamma(2d)}=\frac{d+n}{d}$, yielding, for any nonzero polynomial $p(t)$ of degree at most $n$,
\begin{align}
 \frac{\mathcal S_{n,p}}{\mathcal S_{0,1}}
 &=\frac{\Gamma(2d+2n)}{\Gamma(2d)}
                   \int_{-1}^1 |p(t)|^2\dd\nu_d(t),\label{eq:polynorm}\\
 Q^\times_n[p;y]
 &=\frac{\pi(d+n)m_d}{\sin\delta}
 \frac{\int_{-1}^1 H(y+\arctanh t)|p(t)|^2\dd\mu_d(t)}
      {\int_{-1}^1 |p(t)|^2\dd\nu_d(t)}.
 \label{eq:polyreadout}
\end{align}

To isolate the pure geometric sampling average from the state-dependent boundary normalization, we multiply $Q_n^\times[p;y]$ by the known, $H$-independent positive conversion factor relating $\dd\nu_d$ to $\dd\mu_d$:
\begin{equation}
 \begin{aligned}
 \mathcal A_n[p;y]
 &:=\frac{\sin\delta}{\pi(d+n)m_d}
   \frac{\int_{-1}^1 |p(t)|^2\dd\nu_d(t)}{\int_{-1}^1 |p(t)|^2\dd\mu_d(t)}
   \,Q_n^\times[p;y]\\
 &\hphantom{:}=\frac{\int_{-1}^1 H(y+\arctanh t)|p(t)|^2\dd\mu_d(t)}
       {\int_{-1}^1 |p(t)|^2\dd\mu_d(t)},
 \qquad \deg p\le n.
 \end{aligned}
 \label{eq:Aaverage}
\end{equation}
Because the prefactor in \eqref{eq:Aaverage} is strictly positive, the unit-weight shape average $\mathcal A_n[p;y]$ has the \emph{exact same sign} as the physical readout $Q_n^\times[p;y]$ (with $t=\tanh(x-y)$). Note that for a fixed polynomial $p(t)$, the right-hand side of \eqref{eq:Aaverage} does not depend explicitly on $n$: the radial factor $\mathsf q^{2n}$ cancels from the shape average, so the subscript $n$ simply records the derivative order of the underlying lifted state $|g_{n,p;y,\delta}\rangle$, restricting the allowed degrees to $\deg p\le n$ and fixing the physical norm \eqref{eq:polynorm}. As immediate checks via \eqref{eq:muPd}, $n=0$ and $p(t)=1$ recover the single-insertion scan $\mathcal A_0[1;y]=G_d(y)$, while at even order $n=2N$ the polynomial $p(t)=(1-t^2)^N$ (lifted monomial multiplier $(4\mathsf u\mathsf v)^N$) yields $\mathcal A_{2N}[(1-t^2)^N;y]=(P_{d+2N}*H)(y)$, physically realizing the even subsequence of the differential hierarchy \eqref{eq:diffhierarchy} using only the original scalar species $\OO_\phi$.

\subsection{Local positivity from the moment-matrix hierarchy}
\label{sec:localtests}
Testing $\mathcal A_n[p;y]\ge 0$ simultaneously across all polynomials $p(t)$ of degree at most $n$ reduces to positive semidefiniteness of a finite Hermitian moment matrix~\cite{Lasserre2011,PolyOpt}. Let $\{\pi_j(t)\}_{j=0}^n$ be the real orthonormal Jacobi polynomials with respect to the response measure $\dd\mu_d(t)$ (with $\pi_0(t)=1$), and define the $(n+1)\times(n+1)$ moment matrix at scan center $y$:
\begin{equation}
 M^{(n)}_{jk}(y)=\int_{-1}^1H(y+\arctanh t)
                    \pi_j(t)\pi_k(t)\dd\mu_d(t),
 \qquad 0\le j,k\le n.
 \label{eq:matrix}
\end{equation}
Expanding $p(t)=\sum_{j=0}^n c_j\pi_j(t)$, the shape average \eqref{eq:Aaverage} becomes the Rayleigh quotient $\mathcal A_n[p;y]=c^\dagger M^{(n)}(y)c/(c^\dagger c)$, whose minimum over $c\ne 0$ is the smallest eigenvalue $\lambda_{\min}(M^{(n)}(y))\ge\operatorname*{ess\,inf}H$ (the essential infimum of $H$). Because every basis state in the lifted derivative family $\{|g_{n,\pi_j;y,\delta}\rangle\}_{j=0}^n$ shares the identical homogeneous degree $n$, the radial $\mathsf q$-integral factors out uniformly as one scalar $C_n$. In terms of the boundary-measure overlap matrix $N^{(n)}$ and $C_n$,
\begin{equation}
 N^{(n)}_{jk}=\int_{-1}^1\pi_j(t)\pi_k(t)\dd\nu_d(t),\qquad
 C_n=\mathcal S_{0,1}\frac{\Gamma(2d+2n)}{\Gamma(2d)},
\end{equation}
the unnormalized physical Gram matrix $S^{(n)}$ and stripped cross-response matrix $R_\times^{(n)}(y)$ on this basis are related directly to $N^{(n)}$ and $M^{(n)}(y)$ by
\begin{equation}
 S^{(n)}=C_n N^{(n)},\qquad
 R_\times^{(n)}(y)=C_n\frac{\pi(d+n)m_d}{\sin\delta}M^{(n)}(y).
 \label{eq:twoMatrices}
\end{equation}
Because $N^{(n)}\succ 0$ is strictly positive definite, congruence by $(S^{(n)})^{-1/2}$ preserves positive semidefiniteness: $(S^{(n)})^{-1/2}R_\times^{(n)}(y)(S^{(n)})^{-1/2}\succeq 0\Longleftrightarrow M^{(n)}(y)\succeq 0$.

Positivity of this moment-matrix hierarchy at a \emph{single} fixed scan center $y\in\mathbb R$ already determines the local sign of any bounded displacement profile $H\in L^\infty(\mathbb R)$ almost everywhere along the horizon. One direction is immediate: if $H(x)\ge 0$ almost everywhere, then $c^\dagger M^{(n)}(y)c=\int_{-1}^1 H(y+\arctanh t)|p(t)|^2\dd\mu_d(t)\ge 0$ for every polynomial $p$, so $M^{(n)}(y)\succeq 0$ for all $n\ge 0$. Conversely, suppose $M^{(n)}(y)\succeq 0$ for all $n\ge 0$, so that $\int_{-1}^1 H(y+\arctanh t)|p(t)|^2\dd\mu_d(t)\ge 0$ holds for every polynomial $p(t)$. Because $\dd\mu_d$ is a finite measure on $[-1,1]$, polynomials are dense in $L^2([-1,1],\dd\mu_d)$ by Weierstrass approximation. For any $g\in L^2(\mu_d)$ and any sequence of polynomials $p_k\to g$ in $L^2(\mu_d)$, the boundedness of $H$ gives
\begin{equation}
 \left|\int_{-1}^1 H(y+\arctanh t)\bigl(|p_k|^2-|g|^2\bigr)\dd\mu_d\right|
 \le\|H\|_\infty\|p_k-g\|_2(\|p_k\|_2+\|g\|_2)
 \to 0,
\end{equation}
so the inequality extends to $\int_{-1}^1 H(y+\arctanh t)|g(t)|^2\dd\mu_d(t)\ge 0$ for all $g\in L^2(\mu_d)$. Taking $g$ to be the indicator function of $\{t\in(-1,1):H(y+\arctanh t)\le -a\}$ for any $a>0$ forces that set to have $\mu_d$-measure zero. Since the density of $\dd\mu_d$ is strictly positive on $(-1,1)$ and $t\mapsto y+\arctanh t$ is a smooth diffeomorphism onto $\mathbb{R}$, we obtain the exact local-positivity equivalence
\begin{equation}
 H\ge0\ \text{almost everywhere}
 \quad\Longleftrightarrow\quad M^{(n)}(y)\succeq0
                         \ \text{for every }n\ge 0,
 \label{eq:matrixcriterion}
\end{equation}
which holds pointwise ($H(x)\ge 0$ for all $x\in\mathbb R$) whenever $H$ is continuous.

Equations~\eqref{eq:AbelSuperposition}, \eqref{eq:polyreadout}, and~\eqref{eq:matrixcriterion} resolve the local-positivity question left open by Paper I. For any fixed derivative order $n$ and polynomial $p$ ($\deg p\le n$), applying the projected finite-superposition inequality \eqref{eq:AbelSuperposition} to a fixed finite-difference stencil $|F_h\rangle$ approximating $|g_{n,p;y,\delta}\rangle$ and only then taking $h\to 0$ inside the smooth cross-channel kernel yields $Q_n^\times[p;y]\ge 0$, hence $M^{(n)}(y)\succeq 0$ for every $n\ge 0$, and therefore $H(x)\ge 0$ almost everywhere by \eqref{eq:matrixcriterion}. In particular, if $H$ were negative on any set of positive measure, Weierstrass approximation already guarantees a violation $M^{(n)}(y)\not\succeq 0$ at some \emph{finite} order $n$, avoiding any need to take $n\to\infty$ alongside the weak-gravity or boost limits. This establishes the local nonnegativity of the leading perturbative QES displacement $H(x)$ from single-species boundary modular positivity. Having obtained this general local sign test for an arbitrary bounded profile $H$, we next apply our modular-tomography dictionary to the physical scalar QES in Section~\ref{sec:qes}, before assembling the full ordered-limit derivation, dynamical-versus-structural scope, saturation rigidity, and finite-boost weak-scattering window in Section~\ref{sec:scope}.

\section{The scalar QES and finite-width bounds}
\label{sec:qes}

Our tomography framework in Sections~\ref{sec:inverse} and~\ref{sec:superposition} treated $H(x)$ as an arbitrary bounded profile and characterized how its local sign is encoded in the cross-channel response. We now specialize to the physical QES profile generated by the light scalar primary $\OO$ in the holographic setup of Section~\ref{sec:setup}. Incorporating the bulk dynamical fact that the scalar generalized-entropy source $p_{\rm ent}(z)$ is strictly positive yields a transparent triple-convolution factorization of the readout, exact area and central-thickness benchmarks, and sharp deconvolution-free pointwise bounds on $H(y)$ from a single finite-width probe.

\subsection{Scalar entropy source and the QES profile}

To provide a concrete bulk input for the tomography factorization and pointwise bounds below, we first recall the scalar bulk-entropy source (shape derivative) and perturbative QES equation of Chen and Wang~\cite{Shape,CW}, whose source normalization was independently verified via a two-variable Selberg integral against the cross-channel scattering source in Paper I~\cite{CompanionPositivity}:
\begin{equation}
 p_{\rm ent}(z)=\lambda^{2\Delta}p_*(z),\qquad
 p_*(z)=\mathcal N_\Delta\frac{z^{2\Delta-1}}{(1+z^2)^{2\Delta+1}},
 \label{eq:matching}
\end{equation}
with normalization coefficient
\begin{equation}
 \mathcal N_\Delta=
 \frac{\sqrt\pi\,\Gamma(2\Delta+1)^2}
 {4^{2\Delta}\Gamma(\Delta)^2\Gamma(2\Delta+3/2)}.
 \label{eq:N}
\end{equation}
In our conformal frame, the linearized QES Jacobi equation~\cite{CW} and its Green function $G_L(z,z')=\tfrac12\min(z,z')^2$ (for $L=-\partial_z(z^{-1}\partial_z)$) read
\begin{equation}
 L\eta=4\GN p_{\rm ent},\qquad
 \eta(z)=4\GN\int_0^\infty G_L(z,z')p_{\rm ent}(z')\dd z',
 \label{eq:QES}
\end{equation}
with boundary conditions $\eta(0)=0$ and boundedness at large $z$. Splitting the Green integral at $z'=z$ gives
\begin{equation}
 \eta_*(z)=2\int_0^z z'^2p_*(z')\dd z'
            +2z^2\int_z^\infty p_*(z')\dd z'.
 \label{eq:splitQES}
\end{equation}
For the scalar source \eqref{eq:matching}, this QES profile takes the closed incomplete-beta form
\begin{equation}
 \eta_*(z)=\mathcal N_\Delta\left\{
 B_{X_z}(\Delta+1,\Delta)
 +z^2\bigl[B(\Delta,\Delta+1)-B_{X_z}(\Delta,\Delta+1)\bigr]\right\},
 \label{eq:closedQES}
\end{equation}
where $X_z=z^2/(1+z^2)$ and $B_X(a,b)=\int_0^X t^{a-1}(1-t)^{b-1}\dd t$ is the incomplete beta function. With this bulk profile from refs.~\cite{Shape,CW} fixed from first principles, we now turn to its boundary modular-tomography structure.

\subsection{Source, gravity, and probe factorization}
\begin{figure}[t]
\centering
\includegraphics[width=0.58\textwidth]{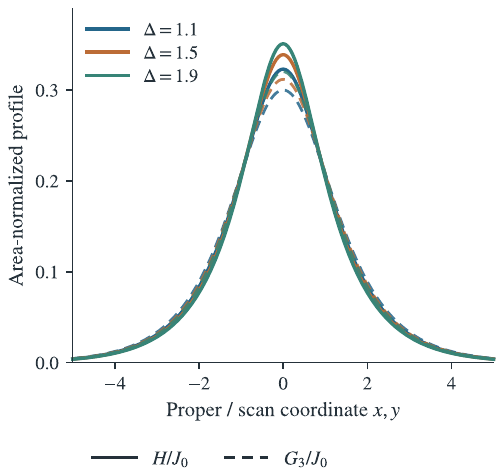}
\caption{The scalar QES and its single-insertion cross-channel readout at probe dimension $d=3$. Solid curves show the area-normalized proper displacement profiles $H(x)/J_0$ and dashed curves show the corresponding single-insertion averages $G_3(y)/J_0$ for $\Delta\in\{1.1,1.5,1.9\}$, where $J_0=\int_{\mathbb R}H(x)\,\dd x$.}
\label{fig:qes}
\end{figure}

Passing to the proper arclength coordinate $x=\log z$ reveals a remarkable structural factorization. Under $\eta_*(z)=e^x H(x)$, the radial Jacobi operator $L=-\partial_z(z^{-1}\partial_z)$ transforms into $L\eta_*=e^{-2x}(1-\partial_x^2)H(x)$. Consequently, the QES Jacobi equation \eqref{eq:QES} becomes a constant-coefficient Helmholtz equation on $\mathbb{R}$,
\begin{equation}
 J(x)=4e^{2x}p_*(e^x)=(1-\partial_x^2)H(x),\qquad
 H=K*J,\quad K(x)=\tfrac12e^{-|x|},
 \label{eq:force}
\end{equation}
where the proper Green function $K(x)=\frac{1}{2}e^{-|x|}$ is selected by the decaying boundary conditions as $x\to\pm\infty$. For the scalar bulk-entropy source $p_*(z)$ in \eqref{eq:matching}, the proper-coordinate source $J(x)$ is proportional to the exact same sech-power family $P_\Delta(x)$ defined in \eqref{eq:convolution}, evaluated at the light scalar dimension $\Delta$:
\begin{equation}
 J=J_0P_\Delta,\qquad
 J_0=2\mathcal N_\Delta B(\Delta+\tfrac12,\Delta+\tfrac12),
 \qquad H=J_0K*P_\Delta.
 \label{eq:scalarH}
\end{equation}
Convolving $H$ with the probe kernel $P_d$ via \eqref{eq:convolution} therefore exhibits the boundary readout $G_d(y)$ as a triple convolution separating the bulk quantum source, transverse gravitational propagation, and finite-width probe averaging:
\begin{equation}
 \underbrace{J_0P_\Delta}_{\text{bulk-entropy source}}
 \ \xrightarrow{\ K\ }\ 
 \underbrace{H}_{\text{proper QES gap}}
 \ \xrightarrow{\ P_d\ }\ 
 \underbrace{G_d}_{\text{boundary readout}},
 \label{eq:sourcegravityprobe}
\end{equation}
or equivalently, in real and Fourier space,
\begin{equation}
 G_d=J_0P_d*K*P_\Delta,\qquad
 \widehat G_d=J_0\frac{T_dT_\Delta}{1+\omega^2}.
 \label{eq:triple}
\end{equation}
We have cross-checked this triple-convolution dictionary numerically across three independent evaluation routes---the coordinate Green integral \eqref{eq:QES}, the incomplete-beta formula \eqref{eq:closedQES}, and direct integration of the scattering source $\Pi_\Delta(z)$ against the probe-convolved Green function---finding agreement to within $9.5\times 10^{-12}$ relative error for $d=3$ and $\Delta\in\{1.1,1.5,1.9\}$ (Figure~\ref{fig:qes}).

\subsection{Absolute area and central-thickness benchmark}
Because both $K(x)$ and $P_\Delta(x)$ are unit-normalized on $\mathbb{R}$, \eqref{eq:scalarH} immediately fixes the total integral $\int_{\mathbb{R}}H(x)\dd x=\int_{\mathbb{R}}G_d(y)\dd y=J_0$. Specializing to the benchmark dimension $\Delta=3/2$, where the incomplete beta functions reduce to elementary rational and inverse-trigonometric functions (Appendix~\ref{app:entropy}), we obtain the exact spatial causal-shadow area and central proper thickness:
\begin{equation}
 J_0=\frac4{35\pi},\qquad
 \mathcal A_{\rm strip}=\frac{6\lambda^3}{35\pi c}+\cdots,
 \qquad H(0)=\frac3{140}-\frac1{35\pi}\approx 0.012334003.
 \label{eq:areanumbers}
\end{equation}
While the integrated causal-shadow area $\mathcal A_{\rm strip}$ is directly accessible from the zeroth moment of a single-insertion scan $G_d(y)$, the local peak thickness $H(0)$ is smeared by the finite probe width $P_d$ unless one either narrows the sampling weight via coherent superpositions (Section~\ref{sec:superposition}) or incorporates bulk positivity of the entropy source, as we do next.

\subsection{Sharp pointwise bounds from a positive entropy source}
When we incorporate the bulk physical prior that the entropy source $J(x)=(1-\partial_x^2)H(x)\ge 0$ is nonnegative, the exponential form of the gravitational Green function $K(x)=\frac{1}{2}e^{-|x|}$ (which obeys $|K'(x)|\le K(x)\le 1/2$) imposes stringent regularity constraints on $H=K*J$:
\begin{equation}
 H\ge0,\quad |H'|\le H,\quad H''\le H,\quad
 H(x)\le\frac12\int H(u)\dd u,
 \label{eq:localbounds}
\end{equation}
and hence, for any nonzero $J\ge 0$, the Harnack-type growth bounds
\begin{equation}
 e^{-|r|}H(x)\le H(x+r)\le e^{|r|}H(x),
 \qquad 0\le z\eta_*'(z)\le2\eta_*(z).
 \label{eq:relativebounds}
\end{equation}
Crucially, this Harnack control allows us to bound the local QES thickness $H(y)$ from above and below using a \emph{single} finite-width measurement $G_d(y)$ at the same point, with no deconvolution required. To derive the sharp constants, consider the even kernel ratio $R_d(u):=(P_d*K)(u)/K(u)$. For $u\ge 0$, splitting the convolution integral at $r=u$ gives
\begin{align}
 R_d(u)&=\int_{-\infty}^u P_d(r)e^r\dd r
           +e^{2u}\int_u^\infty P_d(r)e^{-r}\dd r,\\
 R_d'(u)&=2e^{2u}\int_u^\infty P_d(r)e^{-r}\dd r>0.
\end{align}
Thus $R_d(u)$ increases strictly monotonically on $[0,\infty)$ from its minimum at $u=0$,
\begin{equation}
 \alpha_d:=R_d(0)=\int P_d(r)e^{-|r|}\dd r
        =\frac{2I_{1/2}(d+1,d)}{m_d},
 \label{eq:alpha}
\end{equation}
(where $I_X(a,b)=B_X(a,b)/B(a,b)$ is the regularized incomplete beta function) to its asymptotic limit $R_d(\infty)=\int_{\mathbb{R}}P_d(r)e^r\dd r=1/m_d$. Multiplying $\alpha_d\le R_d(x-y)\le 1/m_d$ by $K(x-y)>0$ and convolving with any nonnegative integrable entropy source $J\in L^1(\mathbb{R})$ ($J\ge 0$, $H=K*J$) immediately gives the pointwise enclosure at every scan point $y\in\mathbb{R}$,
\begin{equation}
 m_dG_d(y)\le H(y)\le\frac{G_d(y)}{\alpha_d}.
 \label{eq:enclosure}
\end{equation}
Both constants $m_d$ and $1/\alpha_d$ are sharp over nonnegative $L^1$ sources: approximating a delta-function source at $x=y$ saturates the upper bound $H(y)/G_d(y)\to 1/\alpha_d$, whereas translating a localized source to $|x-y|\to\infty$ saturates the lower bound $H(y)/G_d(y)\to m_d$.

At $d=3$, \eqref{eq:enclosure} yields the explicit, deconvolution-free sandwich
\begin{equation}
 \frac{75\pi}{256}G_3(y)\le H(y)
                         \le\frac{75\pi}{176}G_3(y),
 \label{eq:d3enclosure}
\end{equation}
where $75\pi/256\approx 0.9204$ and $75\pi/176\approx 1.3387$, and in the presence of bounded measurement noise $|G_d^{\rm obs}(y)-G_d(y)|\le\varepsilon$,
\begin{equation}
 \max\{0,m_d(G_d^{\rm obs}-\varepsilon)\}
 \le H(y)\le (G_d^{\rm obs}+\varepsilon)/\alpha_d.
 \label{eq:noisyenclosure}
\end{equation}

\FloatBarrier

\section{Boundary implications, saturation and the finite boost window}
\label{sec:scope}

The polynomial construction turns positivity of the projected modular response into a local sign constraint. We now derive this consequence and its saturation condition, then analyze the finite-coupling weak-scattering boost window.

\subsection{Local positivity from the modular inequality}
\label{sec:localpos}

Combining the projected finite-superposition boundary inequality \eqref{eq:AbelSuperposition} of Section~\ref{sec:AAinput} with the polynomial moment hierarchy \eqref{eq:matrixcriterion} converts boundary relative-entropy monotonicity into a local sign constraint on any bounded displacement profile $H\in L^\infty(\mathbb R)$ obeying the holographic readout of Section~\ref{sec:superposition}. To see how this works without requiring bounds that are uniform in the derivative order $n$, we proceed in three ordered steps: (i) fix any polynomial degree $n\ge 0$, any nonzero polynomial $p(t)$ of degree at most $n$, and a finite-difference step $h\ne 0$ sufficiently small that every point of the stencil lies at positive Euclidean preparation time and spatial coordinate, letting $|F_h\rangle$ denote the resulting finite superposition of single-insertion Euclidean states approximating $|g_{n,p;y,\delta}\rangle$; (ii) because $|F_h\rangle$ lies in a fixed finite-dimensional subspace of single-insertion preparations, apply the ordered small-interval ($\lambda\to 0^+$), weak-gravity ($\GN\to 0^+$), and positive Laplace ($\epsilon_A\downarrow 0$) limits of Section~\ref{sec:AAinput} directly to $|F_h\rangle$, identifying $\mathfrak R[F_h/\norm{F_h}]$ with the normalized cross-channel quadratic form $Q^\times[F_h/\norm{F_h}]\ge 0$; and (iii) send $h\to 0$ afterwards inside the explicit, smooth off-diagonal cross kernel \eqref{eq:offdiagflux} and boundary Gram norm \eqref{eq:S}, reproducing the polynomial readout $Q_n^\times[p;y]\ge 0$ in \eqref{eq:polyreadout}, and hence $\mathcal A_n[p;y]\ge 0$. Since $n$ and $p$ are arbitrary, every moment matrix obeys $M^{(n)}(y)\succeq 0$, and the completeness relation \eqref{eq:matrixcriterion} yields
\begin{equation}
 H(x)\ge 0\ \text{almost everywhere}
 \qquad(\text{and pointwise if }H\text{ is continuous}).
 \label{eq:localpositivity}
\end{equation}

This order of limits avoids both uniform large-$n$ bounds and any need to differentiate the unprojected estimates of Section~\ref{sec:AAinput} and Appendix~\ref{app:offdiag}: all perturbative and boost limits are taken first on a fixed finite-dimensional single-insertion subspace at fixed $(n,h)$, and $h\to 0$ is taken only inside the resulting smooth cross-channel integral. In particular, if $H$ is negative on any set of positive measure---or is continuous and negative at a single point---Weierstrass density in Section~\ref{sec:localtests} already yields a single fixed polynomial $p$ of finite degree $n$ with $\mathcal A_n[p;y]<0$, producing a contradiction at that fixed finite order $n$. We reiterate the distinction between structural completeness and dynamical input: the local sign constraint \eqref{eq:localpositivity} applies to any bounded profile $H\in L^\infty(\mathbb R)$ whose boundary readout obeys \eqref{eq:AbelSuperposition}; Paper I together with our off-diagonal extension in Section~\ref{sec:AAinput} and Appendix~\ref{app:offdiag} establishes that dynamical dictionary explicitly for the minimally coupled two-scalar Einstein model (where the bulk calculation already gives $H=J_0K*P_\Delta>0$), so \eqref{eq:localpositivity} closes the boundary-to-bulk positivity loop in that model while showing that the same probe hierarchy suffices whenever the linear eikonal readout holds.

\subsection{Saturation}
Once $H(x)\ge 0$ a.e.\ is established, the strict positivity of the single-insertion kernel $P_d(y-x)>0$ on all of $\mathbb{R}$ enforces an immediate rigidity property:
\begin{equation}
 G_d(y_0)=0\text{ for one finite }y_0
 \quad\Longleftrightarrow\quad H=0\text{ almost everywhere}.
 \label{eq:saturation}
\end{equation}
Indeed, if $H\ge 0$ a.e.\ and $\int_{\mathbb{R}}P_d(y_0-x)H(x)\dd x=0$ at even a single scale $y_0$, then $H(x)=0$ a.e.\ on $\mathbb{R}$ (and $H\equiv 0$ everywhere if $H$ is continuous); the same holds if $\mathcal A_n[p;y_0]=0$ for any nonzero polynomial $p(t)$, since $p(t)$ has at most $n$ isolated zeros. We emphasize that \eqref{eq:saturation} characterizes vanishing of the leading $\GN\lambda^{2\Delta}$ QES displacement on the selected branch, rather than exact saturation of the full nonperturbative relative entropy $\PP[F]$ at finite $\GN$ and $\lambda$.

\subsection{Finite boost window and weak-scattering regime}
\label{sec:finitewindow}
While the projected derivation above extracts the tree-level Einstein coefficient prior to taking $\beta\to\infty$, any implementation at finite central charge $c$ (finite $\GN=3/(2c)$) must keep the boosted probe energy below the multi-graviton eikonal threshold. For a single-insertion probe $|f_{s,\delta}\rangle$ of mean boundary energy $E_f=d/\tau$ (with $\tau=s\sin\delta$ and total $v$-directed horizon flux $E_f/2$ from \eqref{eq:flux}), remaining in the weak-scattering regime requires the dimensionless eikonal coupling $g$ to remain small:
\begin{equation}
 g=G_Ne^\beta\frac{E_f}{2}=\frac{G_Ne^\beta d}{2\tau}\ll 1,
 \label{eq:weak}
\end{equation}
giving the necessary Regge boost window
\begin{equation}
 1\ll e^\beta\ll\frac{2s\sin\delta}{G_Nd}.
 \label{eq:single-insertionwindow}
\end{equation}
More generally, on any fixed finite superposition subspace spanned by $\{|f_i\rangle\}_{i=1}^{N_f}$ with Gram matrix $S_{ij}=\langle f_i|f_j\rangle$, destructive interference between non-orthogonal basis states can raise the normalized horizon null flux above the single-insertion maximum $\max_i(E_{f_i}/2)$. Integrating the off-diagonal flux kernel \eqref{eq:offdiagflux} defines the Hermitian positive definite null-momentum matrix
\begin{equation}
 (\mathcal E_+)_{ij}:=\int_0^\infty t_{ij}(z)\,\dd z
 =\frac{d\,S_{ij}}{\bar w_i+w_j},
 \qquad w_i=\tau_i+i\xi_i,
 \label{eq:fluxmatrix}
\end{equation}
so that any normalized superposition $|F\rangle/\norm F$ carries selected null flux $\mathcal E_+[F/\norm F]=(c^\dagger\mathcal E_+ c)/(c^\dagger Sc)\le\mathcal E_{+,\max}:=\|S^{-1/2}\mathcal E_+ S^{-1/2}\|_{\rm op}$, and the mean-flux weak-scattering condition across the subspace is $g_{\max}:=G_Ne^\beta\mathcal E_{+,\max}\ll 1$ (reducing to \eqref{eq:weak} for $N_f=1$).

To compare the coefficient-first projection with a finite-coupling calculation, we use the bounds \eqref{eq:AAprobeBound} and \eqref{eq:crossRemainderBoundII} on the two remainder contributions to the boost-rescaled response. Consider the shifted-window positive average of the extracted coefficient $p_F(\beta)$ over a finite boost interval $[\beta_{\min},\beta_{\max}]$ satisfying the weak-scattering condition:
\begin{equation}
 \widetilde{\mathcal R}_{\beta_{\min},\beta_{\max}}[F]
 =\frac{1}{\beta_{\max}-\beta_{\min}}
  \int_{\beta_{\min}}^{\beta_{\max}} e^{-\beta}p_F(\beta)\dd\beta\ge 0.
 \label{eq:finiteAbel}
\end{equation}
Unlike an unshifted truncated Laplace integral $\epsilon_A\int_0^{\beta_{\max}}e^{-(1+\epsilon_A)\beta}p_F(\beta)\dd\beta$ starting at $\beta=0$ (which suffers a $1-e^{-\epsilon_A\beta_{\max}}$ signal bias and suppresses the $\beta=O(1)$ contamination only logarithmically as $O(L_c^{-1/2})$ for $\epsilon_A=L_c^{-1/2}$), the shifted positive average \eqref{eq:finiteAbel} preserves the growing cross-channel signal $Q^\times[F]$ with \emph{exact unit weight $1$} and discards the $\beta<\beta_{\min}$ region entirely. Choosing the joint scaling window as $\GN\to 0^+$ by
\begin{equation}
 L_c=\log(1/G_N),\qquad
 \beta_{\min}=b_c L_c,\qquad
 \beta_{\max}=a_c L_c,\qquad
 0<b_c<a_c<1,
 \label{eq:jointwindow}
\end{equation}
keeps the eikonal coupling $g_{\max}(\beta_{\max})=\GN^{1-a_c}\mathcal E_{+,\max}\to 0$ strictly inside the weak-scattering regime~\cite{EikonalAdS,SS}, while the exponential remainder bounds \eqref{eq:AAprobeBound} and \eqref{eq:crossRemainderBoundII} yield strict \emph{power-law suppression in $G_N$} of both remainders:
\begin{align}
 \frac{1}{\beta_{\max}-\beta_{\min}}
 \int_{\beta_{\min}}^{\beta_{\max}}
 \big\|S^{-1/2}\mathsf a_{AA}(\beta)S^{-1/2}\big\|_{\rm op}\,e^{-\beta}\dd\beta
 &\le \frac{C_{AA}G_N^{2b_c}}{2(a_c-b_c)L_c},\nonumber\\
 \frac{1}{\beta_{\max}-\beta_{\min}}
 \int_{\beta_{\min}}^{\beta_{\max}} h_\times(\beta)\dd\beta
 &\le \frac{C_\times G_N^{b_c}}{(a_c-b_c)L_c}.
 \label{eq:shiftedWindowErrors}
\end{align}
Thus the shifted-window projection is directly compatible with the coefficient-first limit and suppresses both tree-level remainders by powers of $G_N$. At finite $G_N$ and finite $\lambda$ on a fixed finite probe subspace, applying the same window to the full finite-coupling observable requires that $\lambda$ be small enough for subleading small-interval OPE channels (most notably the $\lambda^4$ stress-tensor channel, which shares the same $e^\beta$ Regge growth, and $\lambda^{2\Delta+2}$ descendants) to remain negligible across $[\beta_{\min},\beta_{\max}]$ and that higher-order multi-graviton corrections remain small across $[\beta_{\min},\beta_{\max}]$. For a discrete finite-difference stencil of step $h$, the two tree-level error bounds in \eqref{eq:shiftedWindowErrors} depend on the actual Gram matrix $S(h)$; if the stencil is varied along with $G_N$, weak scattering and remainder suppression require $G_N^{1-a_c}\mathcal E_{+,\max}(h)\to 0$ together with the sufficient conditions
\begin{equation}
 \frac{N_f\max_{ij} C_{\times,ij}}{s_{\min}(S(h))}
       \frac{G_N^{b_c}}{L_c}\longrightarrow 0,
 \qquad
 \frac{N_f\max_{ij} C_{AA,ij}}{s_{\min}(S(h))}
       \frac{G_N^{2b_c}}{L_c}\longrightarrow 0,
 \label{eq:windowstencil}
\end{equation}
where $C_{AA,ij}=2B(2\Delta,2\Delta)C_{ij}$. For any fixed stencil these conditions follow automatically from $s_{\min}(S)>0$, whereas a derivative evaluated directly on a smooth preparation family uses the fixed-order Cauchy estimates of Appendix~\ref{app:offdiag}.

\section{Discussion}
\label{sec:discussion}

The two papers in this series connect the boundary inequality motivated by Chen and Wang~\cite{CW} to the local geometry of a perturbative quantum causal shadow. Paper I~\cite{CompanionPositivity} establishes the single-insertion boundary observable, matches its growing cross channel non-circularly to the QES displacement, and proves exponential boost suppression of both the complete $AA$ contribution and the boost-rescaled cross-channel Regge remainder to obtain an all-scale averaged positivity inequality. Here we have shown that while single-insertion scale scans uniquely determine $H$ and measure the leading spatial area of the causal shadow without deconvolution, their fixed transverse width can conceal a localized negative displacement, as demonstrated analytically by \eqref{eq:notch}--\eqref{eq:single-insertionpositive}. Upgrading the observable to finite coherent superpositions of the same scalar species---and extending Paper I's tree-level $AA$ and cross-channel remainder bounds to off-diagonal preparation pairs---yields a positive geometric quadratic form whose finite preparation derivatives generate sampling weights proportional to $|p|^2$ for arbitrary polynomials $p$. Because the resulting moment hierarchy is complete for bounded profiles, boundary relative-entropy positivity constrains $H$ locally within the specified Einstein regime, rather than only through broad averages.

The distinction between the physical boundary norm $\dd\nu_d$ and the horizon response measure $\dd\mu_d$ is central to this construction. Both are fixed by the same Euclidean preparation, but they weight the polynomial multiplier differently; retaining both ensures that every polynomial test weight corresponds to a normalized state of the original scalar species and keeps positive semidefiniteness invariant under congruence by $S^{-1/2}$. In the ordered limit, Weierstrass density transfers positivity of all finite moment matrices at a single scan center directly to $H(x)\ge 0$ a.e., enforcing the saturation rigidity \eqref{eq:saturation}, while the shifted-window projection \eqref{eq:finiteAbel}--\eqref{eq:shiftedWindowErrors} converts the exponential remainder bounds into power-law $G_N$-suppression inside the weak-scattering boost regime.

For the scalar QES, the entropy source, gravitational propagation, and probe averaging factorize into a triple convolution. This supplies exact area and central-thickness benchmarks and sharp deconvolution-free pointwise enclosures when the entropy source is positive. Such bulk prior information makes a single finite-width scan directly informative about the local thickness. Natural next steps include extending the dynamical dictionary of Paper I to broader bulk matter sectors and higher-derivative or finite-$\Delta_{\rm gap}$ interactions, and generalizing the tomography to several transverse directions or to backgrounds in which both the causal surface and the QES move.

\section*{Acknowledgments}
L.C. is supported by the
Shuimu Tsinghua Scholar Program of Tsinghua University and the China
Postdoctoral Science Foundation under Grant No.~2025M783401. In the core part of this work, AI assistance was provided primarily by OpenAI's GPT-5.6 Sol, while OpenAI's subsequent GPT-6 to 6.2 Astra/Sol series and Google's Gemini 4-Neon played a major role in manuscript preparation, cross-validation, and further refinements. The author takes responsibility for the manuscript.

\appendix
\section{The absolute scalar benchmark}
\label{app:entropy}

At $\Delta=3/2$, the incomplete beta functions in \eqref{eq:closedQES} simplify to elementary functions. Setting
\begin{equation}
 A_0=\frac{9\lambda^3}{280c},\qquad
 p_{\rm ent}(z)=\frac{12\lambda^3}{35\pi}\frac{z^2}{(1+z^2)^4},
\end{equation}
the coordinate QES profile \eqref{eq:QES} takes the closed elementary form
\begin{equation}
 \frac{\eta(z)}{A_0}
 =z^2+\frac{2(1-z^2)}\pi\arctan z
 -\frac{2z(3+2z^2+3z^4)}{3\pi(1+z^2)^2}.
 \label{eq:rationalQES}
\end{equation}
Since $A_0=\frac{3}{140}\GN\lambda^3$ (using $\GN=3/(2c)$), evaluating \eqref{eq:rationalQES} at $z=1$ ($x=\log z=0$) immediately gives the central proper thickness $H(0)=\eta_*(1)=\frac{3}{140}\bigl(1-\frac{4}{3\pi}\bigr)=\frac{3}{140}-\frac{1}{35\pi}$, while \eqref{eq:scalarH} with $\mathcal N_{3/2}=\frac{12}{35\pi}$ and $B(2,2)=\frac{1}{6}$ yields $J_0=\frac{4}{35\pi}$ and $\mathcal A_{\rm strip}=\GN\lambda^3 J_0=\frac{6\lambda^3}{35\pi c}$, confirming \eqref{eq:areanumbers}. Integrating \eqref{eq:rationalQES} against the $d=3$, $s=1$ radial flux profile $q_3(z;1)=120z^5/(1+z^2)^7$ (via $z=\tan\theta$) gives the exact rational benchmark
\begin{equation}
 \frac1{A_0}\int_0^\infty q_3(z;1)\eta(z)\dd z=\frac{499}{1024},
 \qquad
 \PP_\times^{\rm grow}=\frac{4491\pi}{286720}
                 \frac{E_f\lambda^3}{c}e^\beta.
 \label{eq:absolutebenchmark}
\end{equation}

\section{Off-diagonal extension of the \texorpdfstring{$AA$}{AA} and cross-channel remainder bounds}
\label{app:offdiag}

In this appendix, we extend Paper I's single-insertion tree-level suppression proofs~\cite[appendices~C and~D]{CompanionPositivity} to general off-diagonal preparation pairs $(s_i,\delta_i)\ne(s_j,\delta_j)$ on a fixed $N_f$-dimensional probe subspace $\{|f_i\rangle\}_{i=1}^{N_f}$, establishing the entrywise bounds \eqref{eq:AAentry}, the operator-norm bounds \eqref{eq:AAprobeBound} and \eqref{eq:crossRemainderBoundII}, and their Cauchy extensions to fixed finite preparation derivatives. Because the ordered conformal chart of Paper I~\cite[appendix~C.2]{CompanionPositivity}, the complete tree-level graviton exchange Mellin amplitude $\mathcal M(s_M,t_M)$~\cite[appendix~D.3]{CompanionPositivity}, and the common small-$L$/large-$L$ operator product expansion (OPE) endpoints are independent of $(i,j)$, we focus specifically on the off-diagonal kinematic modifications summarized in Table~\ref{tab:offdiagroadmap}: distinct bra/ket angles $\delta_i\ne\delta_j$, unequal probe radii $(s_i,s_j)$, and the off-diagonal cross-channel phase $\arg(\mathfrak a_{ij}\mathfrak b_{ij})=\delta_i-\delta_j$.

\begin{table}[htbp]
\centering
\small
\begingroup
\renewcommand{\arraystretch}{1.22}
\begin{tabularx}{\linewidth}{@{}>{\raggedright\arraybackslash}p{0.25\linewidth}>{\raggedright\arraybackslash}p{0.23\linewidth}Y@{}}
\toprule
\textbf{Component} & \textbf{Off-diagonal input} & \textbf{Modification relative to Paper I}\\
\midrule
KMS thermal contours (App.~\ref{app:offdiag-kms}) & Distinct angles $\delta_i,\delta_j$ & Joint $i0$ fixes both $AA$ boundary values; a finite probe basis admits common positive thermal gaps.\\
\addlinespace[2pt]
$AA$ global graph envelope (App.~\ref{app:offdiag-edges}) & Probe span $\sigma_p=\delta_i+\delta_j$, radii $(s_i,s_j)$ & Ordered conformal chart maps both words into the open Wightman tube across the interior, one-leg edges, and both mixed corners; light-pair OPE at $L\to 0$.\\
\addlinespace[2pt]
Cross outer region \& small-cross-ratio Mellin (App.~\ref{app:offdiag-cross}) & Complex $\mathfrak a_{ij},\mathfrak b_{ij}$ & Ordered conformal maps control outer endpoints; off-diagonal phases retain strict Stirling margins in the subtracted Barnes integral and analytic continuation fixes the match.\\
\addlinespace[2pt]
Operator norms \& derivatives (App.~\ref{app:offdiag-norms}) & Gram matrix $S\succ 0$, smallest eigenvalue $s_{\min}$ & $\|S^{-1/2}MS^{-1/2}\|_{\rm op}\le s_{\min}^{-1}N_f\max_{ij}|M_{ij}|$; open margins give Cauchy polydisk bounds for fixed derivative orders.\\
\bottomrule
\end{tabularx}
\endgroup
\caption{Summary of the off-diagonal kinematic modifications extending Paper I's complete-graph $AA$ and cross-channel remainder proofs to a fixed $N_f$-dimensional probe subspace.}
\label{tab:offdiagroadmap}
\end{table}

\subsection{Off-diagonal continuation and KMS contours}
\label{app:offdiag-kms}
For a pair of Euclidean preparations $|f_i\rangle$ and $|f_j\rangle$ with polar parameters $(s_i,\delta_i)$ and $(s_j,\delta_j)$ ($\delta_i,\delta_j\in(0,\pi/2)$), the auxiliary real-support Lorentzian correlator $\mathcal M_\PP=\langle[K_A-K_R,\Phi_L]\Phi_R\rangle_\Omega$ of Paper I is continued to the boosted off-diagonal matrix element $\langle f_i|U^\dagger(\beta)\widehat\PP U(\beta)|f_j\rangle$ via Rindler coordinates $(u_R,v_R)=(s_j e^{r_j},-s_j e^{-r_j})$, $(u_L,v_L)=(-s_i e^{-\ell_i},s_i e^{\ell_i})$ along~\cite{CFL}
\begin{equation}
 r_j=\beta+i\delta_j,\qquad
 \ell_i=-\beta-i\pi+i\delta_i,
 \label{eq:offdiagACpath}
\end{equation}
parameterized, up to the initial Wightman shifts specified below, by $h\in[0,1]$ with $\alpha_L=h(\pi-\delta_i)$ and $\theta_R=h\delta_j$ (normalizing by $\sqrt{Z(\tau_i)Z(\tau_j)}=(2\tau_i)^{-d}(2\tau_j)^{-d}$ reproduces the physical overlap factor $S_{ij}=\langle f_i|f_j\rangle$ multiplying the normalized connected four-point function $F_{\rm word}$).

In the right-wedge thermofield trace $\operatorname{Tr}(\varrho^{1/2}X_L^T\varrho^{1/2}Y_R)$, the imaginary right-Rindler Kubo--Martin--Schwinger (KMS) thermal slots take the same structural order as in Paper I~\cite[appendices~C and~D]{CompanionPositivity}:
\begin{itemize}
 \item \emph{Same-side $AA$ channel}: in Paper I's modular coordinates $(j,k)$ for the outside and sandwich orderings ($W_{\rm out}=\langle\OO_g\OO_a\Phi_L\Phi_R\rangle_{\rm conn}$ and $W_{\rm sand}=\langle\Phi_L\OO_g\OO_a\Phi_R\rangle_{\rm conn}$), choosing fixed positive margins $\eta_K,\kappa>0$ satisfying $\eta_K+\kappa<\tfrac{1}{3}\min_i\delta_i$ and displacing $k=y+i\kappa$, $j=x+i(\alpha_L+\eta_K)$ ($\mathrm{out}$) or $j=x+i(\alpha_L-\eta_K-\kappa)$ ($\mathrm{sand}$) keeps all four cyclic thermal gaps strictly positive throughout $h\in[0,1]$ for every pair $(i,j)$:
 \begin{align*}
  \mathrm{out}:&\quad \eta_K,\ \kappa,\ \pi+\theta_R-\alpha_L-\eta_K-\kappa,\ \pi+\alpha_L-\theta_R,\\
  \mathrm{sand}:&\quad \kappa,\ \eta_K,\ \pi+\theta_R-\alpha_L,\ \pi+\alpha_L-\theta_R-\eta_K-\kappa,
 \end{align*}
 since $\pi+\theta_R-\alpha_L\ge2\min_i\delta_i$ and $\pi+\alpha_L-\theta_R\ge\pi$ on the unregulated external path (and interpolating from infinitesimal initial shifts alters these bounds only by $O(\epsilon_p)$). At the real-support end ($h=0$), we set $\Im j=0$, $\Im k=\varepsilon$ with $0<\varepsilon\ll\epsilon_p\ll 1$, assign $\alpha_{L,0}=-2\epsilon_p$ ($\mathrm{out}$) or $+2\epsilon_p$ ($\mathrm{sand}$), and set $\theta_{R,0}=-C\epsilon_p$ for any fixed $C>2$. After transposing the left factors, the two light insertions occupy the adjacent slots $\Im j-\pi$ and $\Im(j+k)-\pi$, following $\alpha_{L,0}-\pi$ in the outside word and preceding it in the sandwich word, so all four starting gaps are strictly positive and the unfolded probe span $\pi+\theta_{R,0}-\alpha_{L,0}$ stays strictly below $\pi$. At these fixed external slots, we linearly interpolate $(\Im j,\Im k)$ to the displaced cycles; smearing the real external supports and taking $\varepsilon\downarrow 0$ followed by $\epsilon_p\downarrow 0$ matches the original Wightman pairing and uniquely fixes the coincidence contact distribution at $k=0$.
 \item \emph{Cross channel}: the imaginary slots in the right-wedge trace are $(\alpha_L-\pi,\Im\sigma-\pi,0,\theta_R)$ for the $\mathrm{TOC}$ word and $(\Im\sigma-\pi,\alpha_L-\pi,0,\theta_R)$ for the $\mathrm{OTOC}$ word. At real supports ($h=0$), we take $\Im\sigma=0$, $\alpha_{L,0}=\mp 2\epsilon_p$ respectively, and $\theta_{R,0}=4\epsilon_p$ (so all four initial gaps are strictly positive), and interpolate linearly to the terminal contours
 \begin{equation}
  \Im\sigma_{\rm TOC}=\pi-\gamma,\qquad
  \Im\sigma_{\rm OTOC}=0,\qquad 0<\gamma<\tfrac12\min_i\delta_i.
  \label{eq:offdiag_theta_bound}
 \end{equation}
 Because all four adjacent gaps are strictly positive at both endpoints, convexity keeps every gap positive along the linear continuation while both contours remain strictly inside the pole-free strip $|\Im\sigma|<\pi$ of the combined modular weight.
\end{itemize}
At fixed external regulators, inserting the entire Gaussian damping factor in the modular times suppresses the vertical closing segments during the initial contour deformations, and carrying the cycles along the continuation to $(\alpha_L,\theta_R)=(\pi-\delta_i,\delta_j)$ preserves thermal ordering just as in Paper I~\cite[appendix~C.3]{CompanionPositivity}; the terminal majorants below then remove the Gaussian regulator at $h=1$ by dominated convergence. As in Paper I~\cite[appendices~C.2--C.3]{CompanionPositivity}, this continuation takes as a perturbative analytic regularity input the standard forward-tube holomorphy and tempered distributional boundary values of the real-time tree-level Einstein Witten diagram under the Skenderis--van Rees prescription~\cite{SkenderisVanRees} (mirroring the CFT forward-tube properties of ref.~\cite{CFTDistributions}). Setting $\rho=e^{-\beta}$ and transferring the external boost $\beta$ at $h=1$ isolates all $\beta$-dependence into the shifted $AA$ kernel $K_\rho(a,g)$ (for $AA$) or into $(u_1,v_1)=(-a\rho,(a\rho)^{-1})$, $(u_2,v_2)=(\rho,-\rho^{-1})$ with $a=e^\sigma$ (for the cross channel), with fixed external probe points $(u_3,v_3)=(L_p,-\bar L_p)$ and $(u_4,v_4)=(R_p,-\bar R_p)$ where $L_p=s_i e^{-i\delta_i}$ and $R_p=s_j e^{i\delta_j}$.

\subsection{Off-diagonal \texorpdfstring{$AA$}{AA} graph envelope via the ordered conformal chart}
\label{app:offdiag-edges}
For a fixed pair $(i,j)$, set $\sigma_p:=\delta_i+\delta_j\in(0,\pi)$ and $L=|a|+|g|$. As in Paper I~\cite[appendix~C.2]{CompanionPositivity}, we bound the complete normalized tree graph directly on the physical Wightman sheet of Appendix~\ref{app:offdiag-kms} across the entire compactified quadrant $(|a|,|g|)\in[0,\infty]^2$.

Passing to the chiral variables $(X_+,X_-)=(u,-1/v)$, a common imaginary Rindler translation centers the two light angles at $-\kappa/2$ and $+\kappa/2$, placing the external probes $\Phi_L,\Phi_R$ at angles $\ell$ and $\ell+\sigma_p$, where $\ell=-(\eta_K+\kappa/2)$ for the outside word and $\ell=+(\eta_K+\kappa/2)$ for the sandwich word, with chiral radii $(s_i,s_i^{-1})$ for $\Phi_L$ and $(s_j,s_j^{-1})$ for $\Phi_R$. As in Paper I~\cite[appendix~C.2]{CompanionPositivity}, we apply the orientation-preserving $\mathrm{PSL}(2,\mathbb R)$ map
\begin{equation}
 m(X)=\frac{1+2X}{1+X},\qquad
 Z=(U,V)=\bigl(m(X_+),-1/m(X_-)\bigr)
\end{equation}
to both chiral lines of the normalized four-point function, yielding physical imaginary parts $r\sin\phi/|1+X|^2$ and $r\sin\phi/|1+2X|^2$ for $X=r e^{i\phi}$. Because $\sigma_p=\delta_i+\delta_j\in(0,\pi)$, choosing $\eta_K$ sufficiently small at the fixed probe parameters $(s_i,s_j,\sigma_p)$ and then $\kappa$ sufficiently small relative to $\eta_K$ keeps $\Phi_R$ bounded below by a strictly positive constant in both components, $\Phi_L$ at magnitude $O(\eta_K)$ with the sign of $\ell$, and both lights at $O(\kappa)$ uniformly over all radii, realizing the strict tube ordering $L,a,g,R$ ($\mathrm{out}$) or $a,g,L,R$ ($\mathrm{sand}$) in both null coordinates. At the Euclidean anchor $|a|=|g|=1$, all four insertions satisfy $v=-u^*$, and Paper I's real conformal homotopy $m_b(X)=b+X/(1+bX)$ ($0\le b\le 1$) preserves these strict inequalities uniformly in $b$ for either probe-radius pair $(s_i,s_j)$, identifying the Wightman sheet throughout the open quadrant.

Every boundary face and the interior of $[0,\infty]^2$ are controlled directly in this chart:
\begin{itemize}
 \item \emph{Both mixed corners ($a\to 0,g\to\infty$ and swap)}: the two light points tend to the real timelike pair $P_0=(1,-1)$ and $P_\infty=(2,-1/2)$ whose difference $(1,1/2)$ has two positive components. Applying a short common complex dilation $Z\mapsto e^{i\omega}Z$ with $|\omega|$ sufficiently small and $\operatorname{sgn}\omega$ chosen so that $\sin\omega\,\Re(Z_g-Z_a)>0$ in both components opens a strictly positive imaginary light--light gap even at the corner while preserving all probe gaps; real dilation invariance continues analytically along this rotation, mapping a closed corner neighborhood into a compact subset of the open ordered tube where $|F_{\rm word}|\le C_{ij}G_N$ for both outside and sandwich words.
 \item \emph{Four one-leg edges and compact interior}: along each one-leg edge, the single finite light radius already retains a strictly nonzero imaginary part in both components of $Z$, keeping all four points inside a compact subset of the open tube (and likewise on every compact subset of $(0,\infty)^2$), yielding $|F_{\rm word}|\le C_{ij}G_N$.
 \item \emph{Common small and large endpoints ($L\to 0$; $A,G\to\infty$)}: as $L\to 0$, the adjacent light insertions fuse near $P_0$ with complex separation $O(L)$ and strictly ordered imaginary gaps bounded below by a positive multiple of $L$, well separated from both probes; the connected light-pair OPE therefore starts at dimension two with double-trace logarithms $O(L^{2\Delta}|\log L|)$, yielding $F_{\rm word}=O(G_N L^2)$ uniformly in $|a|/L\in(0,1)$, while reciprocal radii yield $F_{\rm word}=O(G_N)$ at the common large-radius corner $P_\infty$.
\end{itemize}
Since the probe basis is finite, common margins $(\eta_K,\kappa)$ can be chosen simultaneously for all $N_f^2$ pairs. Patching these open covers gives the global graph envelope $|F_{\rm word}(a,g)|\le C_{ij}\GN\min(L^2,1)$ for every pair $(i,j)$. Feeding this envelope into Paper I's polar-simplex integral~\cite[section~5.2]{CompanionPositivity} and removing the Gaussian regulator by dominated convergence yields the entrywise bound \eqref{eq:AAentry}.

\subsection{Off-diagonal cross-channel remainder}
\label{app:offdiag-cross}
For the cross-channel commutator $-\langle[\delta K_\times,\Phi_L]\Phi_R\rangle_\Omega$, the exact $t$-channel cross ratios at $h=1$ with $L_p=s_i e^{-i\delta_i}$ and $R_p=s_j e^{i\delta_j}$ take the same algebraic form as in Paper I~\cite[appendix~D]{CompanionPositivity},
\begin{equation}
 z_\times(\sigma,\beta)=\frac{(1+a)\rho(R_p-L_p)}{(L_p+a\rho)(R_p-\rho)},
 \qquad
 w_\times(\sigma,\beta)=\frac{(1+a)\rho(\bar L_p-\bar R_p)}
 {(1+a\rho\bar L_p)(1-\rho\bar R_p)},
 \label{eq:offdiag_cross_zw}
\end{equation}
with $\mathsf X=z_\times w_\times$ and $\mathsf Y=(1-z_\times)(1-w_\times)$, and the combined modular weight multiplying $S_{ij}F_\times^{\rm word}(\sigma,\beta)$ is $\mathcal W_\times(\sigma)=[2\cosh(\sigma/2)]^{-(4\Delta+2)}$. We split the $\sigma$-contour at a small basis-dependent threshold $\delta_*>0$.

\paragraph{Outer region $(1+|a|)\rho\ge\delta_*$.}
Denote the boosted left-light radius by $k:=|a|\rho$ (distinct from the $AA$ modular time $k$). On the $\mathrm{TOC}$ contour $\Im\sigma_{\rm TOC}=\pi-\gamma$, the two chiral light coordinates are $k e^{-i\gamma}$ and $\rho$. A common imaginary Rindler translation by $\gamma/2$ shifts the four angles in right-wedge trace order $\Phi_L,\OO_A,\OO_B,\Phi_R$ to $-\delta_i+\gamma/2,-\gamma/2,+\gamma/2,\delta_j+\gamma/2$. Because $\delta_i,\delta_j\in(0,\pi/2)$ and $0<\gamma<\frac{1}{2}\min_i\delta_i$, applying the map $m(X)=(1+2X)/(1+X)$ of Appendix~\ref{app:offdiag-edges} yields strict imaginary gaps in both null coordinates uniformly across the outer region up to all one-leg edges, with the light--light gap closing only at $(k,\rho)=(\infty,0)$, where the same small common dilation $Z\mapsto e^{-i\omega}Z$ ($0<\omega\ll 1$) opens both components of $\Im(Z_B-Z_A)$. (The common small endpoint is excluded by $k+\rho\ge\delta_*$ and the common large endpoint by $\rho\le 1$.) The Euclidean anchor at $k=\rho=1$ and the ordered real homotopy $m_b$ fix the physical $\mathrm{TOC}$ branch.

On the real $\mathrm{OTOC}$ contour $\Im\sigma_{\rm OTOC}=0$, the two light coordinates lie on opposite real wedges at $E=-k\le 0$ and $P=\rho\in[0,1]$, while $\Phi_L$ and $\Phi_R$ already occupy the lower and upper imaginary half-planes in both null coordinates. Fixing any $K>1$, we apply the real $\mathrm{PSL}(2,\mathbb R)$ map $f_C(X)=C+X/(1-X/K)$ on both chiral lines, covering the outer region by the two patches $k\ge\delta_*/2$ (with $C>0$ small) and $\rho\ge\delta_*/2$ (with $C<0$ small in magnitude). On each patch, $f_C(-k)<0<f_C(\rho)$ holds with uniform margins up to the compactified endpoints, so the transformed light points $\bigl(f_C(X),-1/f_C(X)\bigr)$ remain spacelike separated with fixed orientation. A small common imaginary boost $(U,V)\mapsto(e^{i\omega}U,e^{-i\omega}V)$ ($0<\omega\ll 1$) opens both light--light imaginary gaps in the physical $\mathrm{OTOC}$ order $\Phi_L,\OO_A,\OO_B,\Phi_R$ without closing either probe gap; as in Paper I~\cite[appendix~D.2]{CompanionPositivity}, real conformal covariance via $f_{bC}$ and continuation from the ordered real boundary identify this sheet. Covering the outer region by compact subsets of the open ordered tube therefore gives $|F_\times^{\rm TOC}|+|F_\times^{\rm OTOC}|\le C_{ij}G_N$ for every pair $(i,j)$.

\paragraph{Small-cross-ratio region $(1+|a|)\rho<\delta_*$.}
Using $w_i=iL_p$, $w_j=i\bar R_p$, $\mathfrak a_{ij}=\bar w_i+w_j=-i(\bar L_p-\bar R_p)$, and $\mathfrak b_{ij}=\bar b_i+b_j=-i(L_p^{-1}-R_p^{-1})$, expanding \eqref{eq:offdiag_cross_zw} uniformly for $(1+|a|)\rho<\delta_*$ gives
\begin{equation}
 z_\times=i(1+a)\rho\,\mathfrak b_{ij}\bigl[1+O((1+|a|)\rho)\bigr],
 \quad
 w_\times=i(1+a)\rho\,\mathfrak a_{ij}\bigl[1+O((1+|a|)\rho)\bigr].
 \label{eq:offdiag_small_zw}
\end{equation}
Here $\Re\mathfrak a_{ij}=\tau_i+\tau_j>0$, $\Re\mathfrak b_{ij}=\tau_i/s_i^2+\tau_j/s_j^2>0$, and $\mathfrak a_{ij}\mathfrak b_{ij}=|\mathfrak a_{ij}|^2/(w_i\bar w_j)$ carries the off-diagonal phase $\arg(\mathfrak a_{ij}\mathfrak b_{ij})=\delta_i-\delta_j\in(-\pi/2,\pi/2)$. On the real $\mathrm{OTOC}$ contour, $\operatorname{Log}_{\rm OTOC}\mathsf Y=\log_{\rm loc}\mathsf Y+2\pi i$ (with $\log_{\rm loc}1=0$), whereas the $\mathrm{TOC}$ word carries zero winding: Paper I fixes these branches at a separated Euclidean anchor, and varying bra and ket along any compact path of physical preparations keeps both chiral ratios inside small punctured disks (since $\Re\mathfrak a_{ij},\Re\mathfrak b_{ij}>0$ and the light insertions stay separated on each contour), preserving the winding identically. Setting $\epsilon_Y=-\log_{\rm loc}\mathsf Y=z_\times+w_\times+O([(1+|a|)\rho]^2)$ after extracting the $\mathrm{OTOC}$ jump, combining \eqref{eq:offdiag_small_zw} with the real $\mathrm{OTOC}$ line $\Im\sigma_{\rm OTOC}=0$ in \eqref{eq:offdiag_theta_bound} (where $a=e^\sigma>0$) yields the uniform phase margins
\begin{equation}
 \Im\epsilon_Y\ge c_0(1+|a|)\rho>0,
 \qquad
 |\arg\mathsf X-\pi|\le\frac{\pi}{2}-c_1,
 \label{eq:offdiag_phase_margins}
\end{equation}
for constants $c_0,c_1>0$ depending only on the finite probe basis.

Abbreviating $A=\mathfrak a_{ij}$, $B=\mathfrak b_{ij}$, and $h_Y=1-\mathsf Y$, the lifted phases on the real $\mathrm{OTOC}$ contour obey $\arg\mathsf X=\pi+\arg A+\arg B+O((1+a)\rho)$ and $\arg(\mathsf X/h_Y^2)=\arg A+\arg B-2\arg(A+B)+O((1+a)\rho)$. Because $\Re A,\Re B>0$, $\arg(A+B)$ lies in the closed interval between $\arg A$ and $\arg B$, so $|\arg A+\arg B-2\arg(A+B)|\le|\arg A-\arg B|<\pi$, supplying a strict common Stirling margin (exceeding $\pi/2$ on the regular branch and $\pi$ on the singular branch) across the finite probe basis. In the summed Mellin representation of Paper I~\cite[appendix~D.4]{CompanionPositivity}, we subtract the leading $h_Y^{-1}$ Barnes contribution before shifting the remaining contour from $\Re v=-1/4$ to $\Re v=+1/4$. At the sole pole $v=0$, the subtracted kernel satisfies $\mathcal T(0,\mathsf Y)=\mathsf Y/h_Y-1/h_Y=-1$, giving a bounded constant residue; on $\Re v=+1/4$, the regular and singular Gauss branches of the subtracted kernel are $O(|\mathsf X|^{1/4})$ and $O(|\mathsf X|^{1/4}|h_Y|^{-1/2})$, respectively. Since $|\mathsf X|\asymp |h_Y|^2\asymp[(1+a)\rho]^2$ for each pair, the subtracted remainder is uniformly $O(1)$ throughout the inner region, while the $\mathrm{TOC}$ direct light-pair OPE yields $O(G_N[(1+|a|)\rho]^2)$. The $\mathrm{OTOC}$ discontinuity therefore isolates the leading growing term
\begin{align}
 \frac{\Delta F^{\rm lead}(\sigma,\beta)}{G_N}
 &=\frac{16\pi\Delta d}{(1+a)\rho}\int_{\Re v=-1/4}\frac{\dd v}{2\pi i}\,
   \mathcal S(2v)\Gamma(\Delta-v)\Gamma(d-v)\Gamma(2v+1)\nonumber\\
 &\qquad\times\frac{(\mathfrak a_{ij}\mathfrak b_{ij})^v}{(\mathfrak a_{ij}+\mathfrak b_{ij})^{2v+1}}
  =-e^\beta\frac{q_{\times,ij}(\sigma)}{\langle\OO_A(\sigma)\OO_B\rangle}.
 \label{eq:offdiag_exact_eikonal}
\end{align}
To verify the final equality for complex $(A,B)=(\mathfrak a_{ij},\mathfrak b_{ij})$, divide the off-diagonal radial QES readout by the overlap $S_{ij}$: comparing \eqref{eq:S} with \eqref{eq:offdiagflux} gives the exact identity $4\pi n_i^*n_j/S_{ij}=2(AB)^d/\Gamma(d)^2$ (with fractional powers continued from real positive $A,B$). The radial integral is holomorphic on the product half-plane $\Re A,\Re B>0$ because $\Re(A+Bz^2)>0$ for all $z>0$, and the Barnes integral in \eqref{eq:offdiag_exact_eikonal} is holomorphic on the same domain by the strict phase margin $|\arg[AB/(A+B)^2]|<\pi$. Because the two holomorphic functions agree identically on the nonempty real open subset $\{(A,B)\in(0,\infty)^2:AB<4\}$ swept out by diagonal preparations via Paper I's Fourier--Barnes match~\cite[appendix~D.4]{CompanionPositivity}, two-variable analytic continuation extends their equality to all complex pairs $(A,B)$ with $\Re A,\Re B>0$. Here $S_{ij}q_{\times,ij}(\sigma)=16\pi\int_0^\infty\!\dd z\,\dd z'\,t_{ij}(z)G_L(z,z')[J_{-\sigma}(z')/(iz')]$ in the original modular variable, so integrating the $(1+a)^{-1}$ factor against the modular weight with the commutator minus sign reproduces the exact off-diagonal matrix element $Q^\times_{ij}$ in \eqref{eq:crossform} with its full interference phase.

In the outer region of the real $\mathrm{OTOC}$ contour, the extended leading integrand is uniformly bounded because $(1+a)\rho\ge\delta_*$, so the subtracted $\mathrm{OTOC}$ integrand is bounded there as well. Because the combined modular weight $\mathcal W_\times(\sigma)$ is integrable along both contours in \eqref{eq:offdiag_theta_bound}, integrating the inner and outer bounds yields $|\mathsf p_{\times,ij}(\beta)-e^\beta Q^\times_{ij}|\le C_{\times,ij}$ at large $\beta$, while local boundedness of the ordered integrals on any compact boost interval extends this to all $\beta\ge 0$, establishing \eqref{eq:crossRemainderBoundII}.

\subsection{Finite-subspace operator norms and preparation derivatives}
\label{app:offdiag-norms}
On any fixed $N_f$-dimensional probe subspace with physical Gram matrix $S\succ 0$ and smallest eigenvalue $s_{\min}>0$, the matrix norm inequality $\|S^{-1/2}M S^{-1/2}\|_{\rm op}\le s_{\min}^{-1}N_f\max_{i,j}|M_{ij}|$ converts the uniform entrywise bounds $|\mathsf a_{AA,ij}(\beta)|\le 2C_{ij}B(2\Delta,2\Delta)e^{-\beta}$ and $|\mathsf r_{\times,ij}(\beta)|\le C_{\times,ij}$ directly into the operator-norm bounds \eqref{eq:AAprobeBound} and \eqref{eq:crossRemainderBoundII}. Finally, because the bra and ket preparation coordinates can be complexified independently while preserving the compactness of the separated endpoint charts as well as the strict angular margins of both the direct light-pair OPE and the two Gauss branches, the integral majorants remain uniform on small preparation polydisks. Cauchy's integral formula therefore bounds every fixed finite order of preparation derivatives by the same $O(e^{-\beta})$ boost decay, with constants depending on the derivative order, polydisk radii, and the derivative-space Gram matrix.

\bibliographystyle{JHEP}
\bibliography{ref}

\end{document}